\documentclass[printer]{aa}

\usepackage{txfonts}

\usepackage{subfig}

\usepackage{epstopdf}
\usepackage{fixltx2e}
\usepackage{float}

\usepackage{lscape}

\usepackage{enumitem}

\usepackage{natbib}

\usepackage{hyperref}

\newcommand{\CaII}{Ca\,{\sc ii} }
\newcommand{\mets}{\,\mathrm{m\,s^{-1}}}

\newcommand{\seconds}{\,\mathrm{s}}
\newcommand{\Mm}{\,\mathrm{Mm}}
\newcommand{\minutes}{\,\mathrm{min}}
\newcommand{\hr}{\,\mathrm{hr}}
\newcommand{\simm}{\mathtt{\sim}}
\newcommand{\arcs}{\,\mathrm{arcsec}}

\newcommand{\MK}{\,\mathrm{MK}}
\newcommand{\px}{\,\mathrm{px}}

\newcommand{\G}{\,\mathrm{G}}
\newcommand{\DNS}{\,\mathrm{DN\,s^{-1}}}
\newcommand{\Mx}{\,\mathrm{Mx}}
\newcommand{\erg}{\,\mathrm{erg}}

\newcommand{\nm}{\,\mathrm{nm}}

\newcommand{\NaI}{Na\,{\sc i} }

\newcommand{\e}[1]{\ensuremath{\times 10^{#1}}}
\newcommand{\esim}[1]{\sim\ensuremath{10^{#1}}}

\newcommand{\Bmdi}{{B_z}_{\,\mathrm{MDI}}}
\newcommand{\Bnfi}{{B_z}_{\,\mathrm{NFI}}}
\newcommand{\VInfi}{{V/I}_{\;\mathrm{NFI}}}
\newcommand{\Wfree}{W_{\mathrm{free}}}

\newcommand{\Fig}[1]{Fig.\,\ref{#1}}
\newcommand{\Figs}[1]{Figs.\,\ref{#1}}
\newcommand{\Eq}[1]{Eq.\,\ref{#1}}

\newcommand{\Section}[1]{Sect.\,\ref{#1}}

\begin{document}

\title{Relationship between supergranulation flows, magnetic cancellation and network flares}

\author{R. Attie\inst{\ref{inst1}}\and D. E. Innes\inst{\ref{inst1}}\and S.K. Solanki\inst{\ref{inst1},\ref{inst2}}\and K.H. Glassmeier\inst{\ref{inst3}}}

\institute{Max-Planck-Institut f\"{u}r Sonnensystemforschung, 37077 G\"ottingen, Germany \label{inst1}
\and School of Space Research, Kyung Hee University, Yongin, Gyeonggi, 446-701, Republic of Korea \label{inst2}
\and Institute of Geophysics and extraterrestrial Physics, Technische Universit\"{a}t Braunschweig, Mendelssohnstrasse 3, D-38106 Braunschweig, Germany}  \label{inst3}

\abstract
{Photospheric flows create a network of often mixed-polarity magnetic field in the quiet Sun, where small-scale eruptions and network flares are commonly seen.}
{The aim of this paper is (1) to describe the characteristics of the flows that lead to these energy releases, (2) to quantify the energy build up due to photospheric flows acting on the magnetic field, and (3) to assess its contribution to the energy of small-scale, short-lived  X-ray flares in the quiet Sun. }
{We used photospheric and X-ray data from the SoHO and Hinode spacecraft combined with tracking algorithms to analyse the evolution of five network flares. The energy of the X-ray emitting thermal plasma is compared with an estimate of the energy built up due to converging and sheared flux.}
{Quiet-Sun network flares occur above sites of converging  opposite-polarity magnetic flux that are often found on the outskirts of network cell junctions, sometimes with observable vortex-like motion. In all studied flares the thermal energy was more than an order of magnitude higher than the magnetic free energy of the converging flux model. The energy in the sheared field  was always higher than in the converging flux but still lower than the thermal energy.}
{X-ray network flares occur at sites of magnetic energy dissipation. The energy is probably built up by supergranular flows causing systematic shearing of the magnetic field. This process appears more efficient near the junction of the network lanes. Since this work relies on 11 case studies, our results call for a follow-up statistical analysis to test our hypothesis throughout the quiet Sun.}


\maketitle

\section{Introduction}

The dynamics of the quiet Sun's upper atmosphere is driven by sub-surface convective motions that entrain small-scale magnetic flux concentrations  and create a network of supergranular cells \citep{Schrijver97, Parnell01, Priest02} outlined by bright chromospheric emission.  The magnetic field appears to be crucial to the heating of the  chromospheric network. Both wave and magnetic dissipation mechanisms could supply  the necessary energy  \citep[e.g.][]{Depontieu07, Hasan08, Balle11, Meyer13, Depontieu14} but at present there is no clear consensus about which dominates.  Above the chromosphere, the slightly hotter transition region network is characterised by small, highly dynamic brightenings and jets \citep{McIntosh07, Aiouaz08} that are thought to result from magnetic reconnection \citep{Detal91,IIAW97b}. At higher  temperatures, the network mixes with diffuse coronal emission and loops \citep{Feldman00} in which sudden small-scale X-ray brightenings are seen at a rate of about one every three seconds for the whole Sun \citep{Krucker97}.

Since the beginning of the SoHO era, it is possible to combine long time series with high resolution, simultaneous and co-spatial analyses from the photosphere up to the low corona. Studies have revealed that supergranular flows acting on magnetic field concentrations may be a possible energy source of X-ray brightenings \citep{Potts07}. The junctions of the cells are sites of vortex-like flows that drag magnetic field  concentrations \citep{Attie09} toward their centre and lead to small-scale (few Mm) CME-like eruptions \citep{Innes09}, and transition region explosive events \citep{Innes2013}.  

This paper provides a first analysis of the role of flows and vortex-like motions in quiet Sun network flares, but further statistical analysis is needed. Here we use a combination of modern algorithms to investigate the relationship between photospheric flows, magnetic field,  and  small-scale heating processes, seen as X-ray transients, in the quiet Sun. Similar X-ray brightenings were shown by \citet{Krucker97}  and \citet{Krucker00} to have flare-like characteristics, and are thus thought to be triggered by the same process as flares: magnetic reconnection. An investigation of the photospheric magnetic field below small, active-region X-ray transients found that half were related to flux emergence, but that there was no obvious flux evolution associated with the others \citep{Shimizu02}. 

 We use high-resolution observations from Hinode \citep{Kosugi07}, and aim at relating the photospheric flows and the evolution of the magnetic flux to the soft X-ray brightenings in the quiet Sun. 
 The flows are derived using the balltracking algorithm \citep{Potts04} which has proved very good at resolving photospheric flows from the Michelson Doppler Imager (MDI) and the solar optical telescope \citep[SOT,][]{Tsuneta08} continuum images  \citep{Attie09}. The evolution of the magnetic field is tracked using magnetic balltracking which is an efficient magnetic flux analysis framework\footnotemark \citep{Attie15} and \citep{Attie2015a}. 
 \footnotetext{The codes used for tracking magnetic elements with magnetic balltracking are available as an open-source project hosted on Bitbucket at \url{https://bitbucket.org/raphaelattie/balltracking-framework}}

The paper is organised as follows: in Sections~\ref{sec:multi_instr_obs} and \ref{sec:protocol} we describe the observations and their co-alignment. The calibration of the magnetograms from the narrow-band filter imager (NFI) on SOT is explained in Section~\ref{sec:NFI_calibration}. In Section~\ref{sec:multi_layer_analyses} we present case studies of small-scale X-ray events in the low corona and the resulting coronal heating. In the final section, we discuss the plausible theoretical implications of this ubiquitous quiet-Sun activity. 

\section{Observations \label{sec:multi_instr_obs}}

Five instruments were involved in co-spatial observations on September 26, 2008: MDI \citep{Setal95} on SoHO, the X-Ray Telescope \citep[XRT,][]{Golub07}, the broad-band filter imager (BFI) and the narrow-band filter imager (NFI) of the solar optical telescope \citep[SOT,][]{Tsuneta08} onboard Hinode. Each of them provided an $8\hr$ time series of data, from 15:00~UT to 23:00~UT. They were pointing near disk centre. 
Images from the full-disk extreme ultraviolet imaging telescope \citep[EIT,][]{Delaboudiniere1995} on SoHO are also used for co-aligning the data. See Table \ref{tableObservations} for more details on the data sets. 
Due to a long data gap in the middle of these observations, this study focuses on the first 4 hr time series. In what follows, the times are given in universal time (UT).

High resolution MDI images provided the magnetic field and continuum images from which photospheric flows were computed over a large field-of-view (FOV). The SOT images covered a smaller FOV at higher resolution, and likewise provided magnetic field and photospheric flows for the region. Sites of coronal heating were revealed by XRT images.

\begin{table*}
\centering
\caption{Summary of the instrument parameters used for our observations.}
\label{tableObservations}
\begin{tabular}{lcccc}
\hline
Instruments  & Product type & Cadence &  FOV ($\arcs^2$) & Pixel size ($\arcs$) \\
\hline
XRT/Hinode  & C-poly ($1\MK < T < 10\MK$) images & 30s & $384 \times 384 $ & 1 \\
\hline
SOT-BFI / Hinode & Blue continuum \& Ca II images & 30s / 90s & $214 \times 212$ & 0.22 \\
\hline
SOT-NFI/ Hinode & Stokes V/I filtergrams & 2 min & $214 \times 212$ & 0.3 \\
\hline
MDI/SoHO & Magnetograms & 1 min & $660 \times 330$ & 0.6 \\
\hline
MDI/SoHO & Continuum & 1 min & $660 \times 330$ & 0.6 \\
\hline
MDI/SoHO & Magnetograms & 90 min & Full disk & 2 \\
\hline
EIT/SoHO & EUV images & 1 image available & Full disk & 2.6 \\
\hline
\end{tabular}

\end{table*}

\begin{figure} %/Users/attie/IDL/routines/main_scripts/08Sep26/overview.pro
	\centering
	\includegraphics[width = 1 \columnwidth]{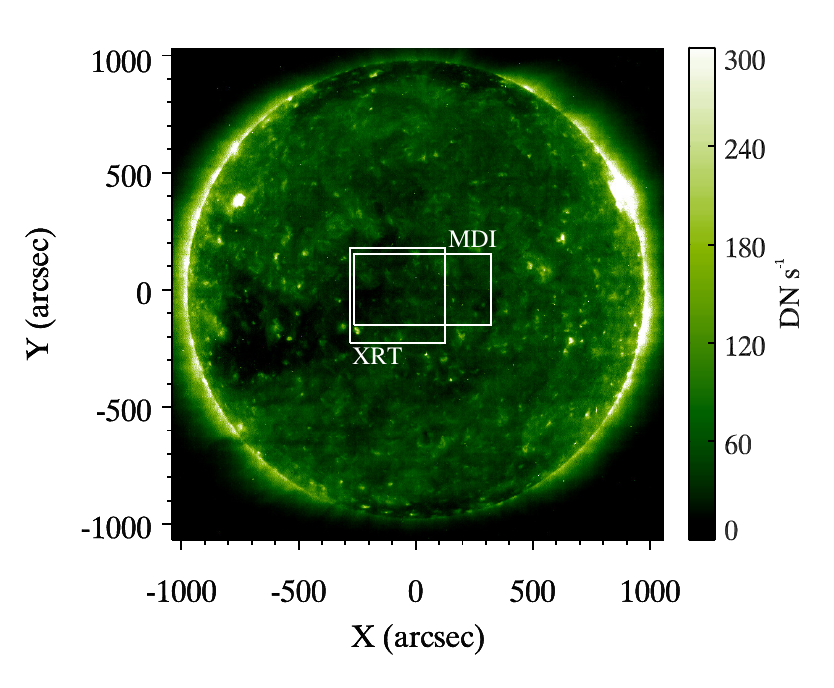}\\
	\includegraphics[width = 1 \columnwidth]{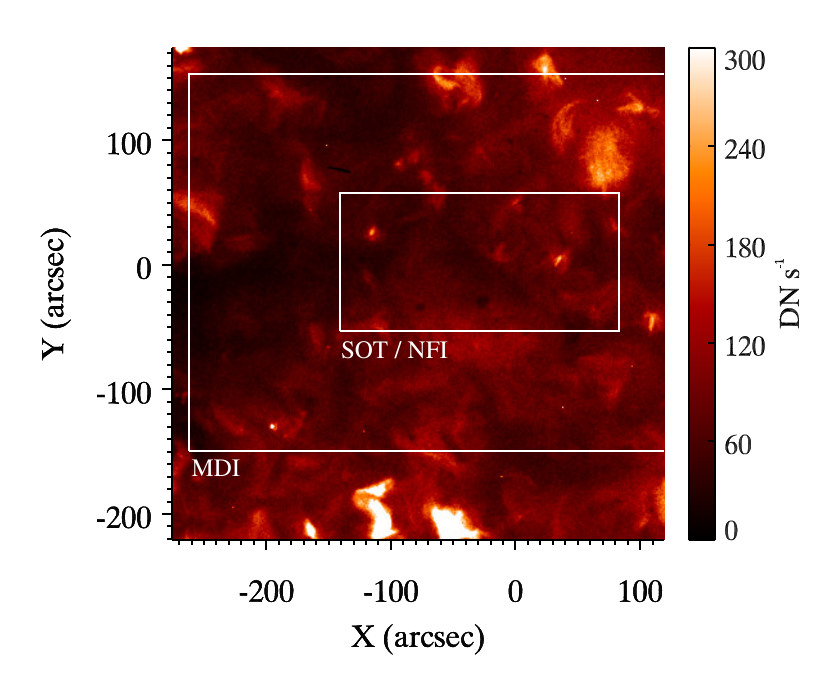}
	\caption{Top: EIT full disk context image  (September 26, 2008 - 19:12 UT). XRT and MDI FOVs are shown in white rectangles. Bottom: XRT FOV at 15:01 UT  with the analysed MDI and NFI/SOT FOVs shown in white rectangles. The width of the MDI FOV is wider than XRT's.} 
	 \label{overview0}
\end{figure}

The regions observed by each instrument selected for analysis are presented in \Fig{overview0}, with (respectively for each figure) EIT and XRT data as the background image. The MDI images were read out from half the full detector. There was no full-resolution EIT image available at the beginning of the time series, so the displayed image is the closest available, at 19:12, in the 195\AA~wavelength.

\section{Co-alignment \label{sec:protocol}}
This study describes small-scale events in both the photosphere and the low-corona that are observed with different instruments. This demands an accurate co-alignment of nearly 1-$\mathrm{arcsec}$. Since MDI has a pointing accuracy below $1\arcs$, it is used as the reference. The EIT full disk image taken at the same time as the MDI full disk (19:12) is also used as a co-alignment frame for X-ray images. Pointing information is given in the FITS headers of the MDI and EIT data, and is used to directly co-align both instruments. Here we give a summary of a long co-alignment procedure. More details and illustrations are available in \citet[][\S~4]{Attie15}.

\subsection{Registration of MDI data \label{sec:mdi_reg}}
As Hinode is on a polar orbit around Earth, and SoHO is on the Lagrange point L1, the MDI data are interpolated to the Earth view by reducing its pixel size by a factor of 1.01.
All images are de-rotated rigidly to the same time using the empirical formula of the solar differential rotation \citep{Howard90} at the local latitude.

\subsection{Registration of BFI/SOT dataset \label{BFI_coalignment} \label{sec:BFI_reg}} 

\begin{figure} %/Users/attie/IDL/routines/coalign/coalign_legend_AA_Paper.pro 
	\centering
	\includegraphics[width = 1\columnwidth]{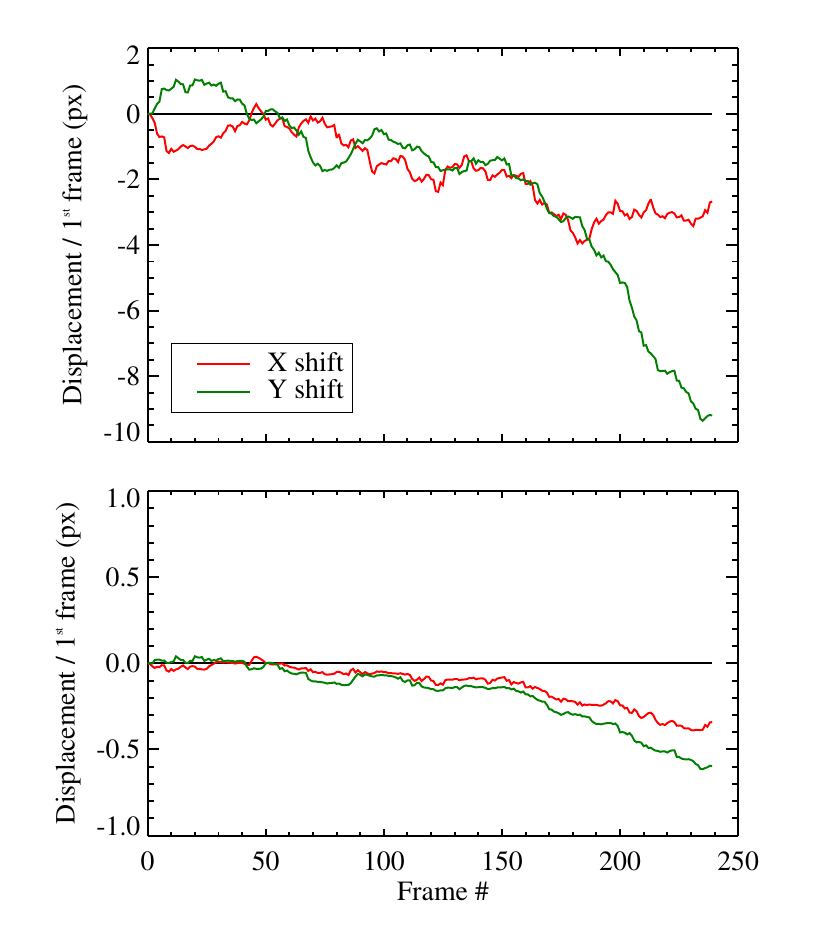}
	\caption{Top: Initial alignment errors between the BFI images, before the co-alignment procedure. Bottom: Alignment offsets after the co-alignment procedure. The shifts are given with respect to the first frame.} 
	\label{BFI_Tracking_Curves}
\end{figure}

The interval between two blue continuum images from BFI changes alternatively from $90\mathrm{s}$ to $30\mathrm{s}$. The registration is done with the Solarsoft routine "fg\_rigidalign.pro" that is dedicated to SOT images. It is not possible to cross-correlate accurately individual images taken more than $3\minutes$ apart because the granulation changes significantly on this timescale. Here, the offsets are calculated by cross-correlating consecutive pairs of images and then shifting the images by their offsets. Even after applying the shifts there are small residual errors. The accumulated cross-correlation shifts (alignment errors) before and after applying the procedures are shown in \Fig{BFI_Tracking_Curves}. Each data point on the curve is the cumulated shift between images $1$ and $2$, images $2$ and $3$, and so on up to image $n$. So the shifts shown at a given image $n$ are the cross-correlation offsets with respect to the first image.
We have a maximum displacement of more than $10\px$ before applying the procedure (top), and about $0.5\px \sim 0.1\arcsec$ after co-alignment (bottom). The small drift after the pairwise cross-correlation, seen in the bottom frame, may be due to a real drift in the granulation or systematic residual errors in the pairwise correlations and is therefore not applied to the data.  The co-aligned time series are finally separated into two series, with a regular time cadence of $2\minutes$, more suitable for balltracking.

\subsection{Co-alignment of NFI/SOT with MDI high-resolution images \label{sec:NFI_reg}} 

NFI shares the same CCD as BFI, so both instruments cannot record data at the exact same time. 
Each NFI frame is co-aligned with the nearest in time MDI high-resolution magnetogram (separated in time by $30\seconds$ at most). We found that apart from vertical and horizontal shifts, the SOT images needed to be rotated by 0.65 degrees.
By visual inspection, the uncertainty of this co-alignment is estimated to be of the order of $\pm1$ NFI pixel ($\pm0.3\arcs$).

\subsection{Co-alignment of NFI magnetograms with the blue continuum images from BFI}

The photospheric granulation seen in the blue continuum with BFI, and the small magnetic elements observed in the NFI magnetograms are, geometrically speaking, completely different features. We use the \CaII BFI images in an intermediate step to coalign the two.
The brightest features in \CaII are geometrically similar to the magnetic features in the NFI filtergrams and the granulation seen in \CaII can be matched with the blue continuum, although in the \CaII line the granulation is seen at heights closer to the temperature minimum, and their brightness appears reversed  \citep{EvansCat72,Suemoto87}. 
\citet{Rutten04} calculate a maximum anti-correlation of 50\% between the \CaII intensity and the photospheric granulation when taking into account a time delay of $2$--$3\minutes$. 
So we use the BFI \CaII images as intermediate co-alignment frames. First we calculate their misalignment with respect to the blue continuum images. Second, we calculate the \CaII image misalignment with respect to the NFI magnetograms. Finally, the shifts between the pairs blue continuum - \CaII, and \CaII - magnetograms (NFI) are summed up to obtain the shifts between the pair blue continuum (BFI) - magnetograms (NFI). Both series of misalignment errors are calculated by cross-correlation; following the same procedure as in \S~\ref{sec:BFI_reg}, and from which we estimate our co-alignment uncertainty to be about $\pm 0.6\arcs$.

\subsection{Co-alignment of the SoHO and Hinode data}
The full-disk, low-resolution MDI magnetogram at 18:59 is used as an intermediate reference map, mapped to the Earth view, and is interpolated to have the same pixel size as the MDI high-resolution magnetogram. The latter, recorded at the same time, is co-aligned to the full disk magnetogram using cross-correlation, with an uncertainty of $\pm 0.3\arcs$.

We estimate the uncertainty of the coordinates of the disk centre of both full disk images of MDI and EIT (SoHO) to be negligible compared to the other alignment uncertainties.
%The alignment of XRT involved the full-disk EIT image at 195~\AA. This wavelength (among 4 different ones available in EIT) offered the best resemblance with the features visible in XRT. Both regions are taken at the same reference time (18:59). 
The XRT image series are registered using cross-correlation. This series is co-aligned to the full-disk EIT image at 195\AA\ by cross-correlation, assisted manually using the similar features present in both datasets. The co-alignment error is within $\pm0.5\arcs$. Next, we used the shifts of the XRT frames with respect to the EIT frame  to co-align the XRT and MDI high-res. data. 

Finally, we have co-aligned the data series within the co-alignment uncertainty given in Table \ref{table_coalignment}. These are random co-alignment errors, hence the co-alignment error between any different pair of instruments is obtained by taking the quadratic sum of the relevant errors in Table \ref{table_coalignment}. 

\begin{table}
\caption{Co-alignment random error between the image series.}
\begin{tabular}{lcccc}
\hline
Instruments  & BFI-NFI & NFI-MDI &  MDI-EIT & EIT-XRT \\
\hline
Error  & 0.6" & 0.3" & 0.3" & 0.5" \\
\hline
\end{tabular}
\label{table_coalignment}
\end{table}

\section{Calibration of the magnetograms from NFI/SOT \label{sec:NFI_calibration}}

\subsection{General approach}
The observations from the narrow-band filter imager (NFI) are the ratio, made onboard, of the Stokes V and Stokes I images from the narrowband filtergrams in the \NaI line at $589.6\nm$ \citep[][\S~5.1]{Tsuneta08}. In our data, they are given in arbitrary units with a polarity opposite to those of the MDI magnetograms (i.e. the positive values in MDI correspond to negative values in the NFI filtergrams, and vice versa). An accurate calibration of the filtergrams into units of magnetic field requires the data from the spectro-polarimeter (SP) from SOT/Hinode which provides line profiles \citep{Chae2007}. Unforturnately, SP observations are not available for the period,  and we have to revert to another method. 
Instead, we convert the V/I filtergrams into physical units (G) using the high-resolution MDI magnetograms (SoHO), in a similar manner to \citet{Parnell08a}. As our original data were given in different arbitrary units than the one in the latter paper, our calibration factor will be different. 
The calibration consists in fitting the NFI data to the MDI data to derive a linear calibration factor that is used to rescale the units of the NFI filtergrams, which forces both data series to have the same flux density. We acknowledge that the magnetograms used here are bidimensional spatial distributions of the approximated line-of-sight magnetic field. Near the disk centre, it is considered parallel to the heliocentric Z-axis (pointing toward the observer), referred to as $B_z$, and indexed by the name of the instrument with which it is measured (e.g. $\Bmdi$, $\Bnfi$). Once calibrated, the NFI filtergrams will also be referred to as "magnetograms".  We are aware that a magnetogram based on a filtergram at a single wavelength suffers from considerable uncertainty, as changes in the line shift or width are misrepresented as changes in $B_z$.

\subsection{Least-squares fits \label{sec:NFI_rejection}}
The NFI uses a CCD that is divided in two parts. The left half of the images was impaired by time-dependent artefacts.
Because this artefact is time-dependent, and stops precisely at the middle of the CCD, we compute the calibration coefficient only for the right-hand side of the NFI filtergrams.
As said earlier, the polarity of the original data are opposite to those of MDI. For simplicity, we first reverse the sign of the NFI data, which will now be referred to as $\VInfi$. Thus the correlations derived below are, in fact, anti-correlation with respect to the original, non-reversed data.

\begin{figure}%/Users/attie/Matlab/raphael/codes/08Sep26/MDI_SOT/calib4_cross_corr_AA_Paper.m
	\centering
	\includegraphics[width=1 \columnwidth]{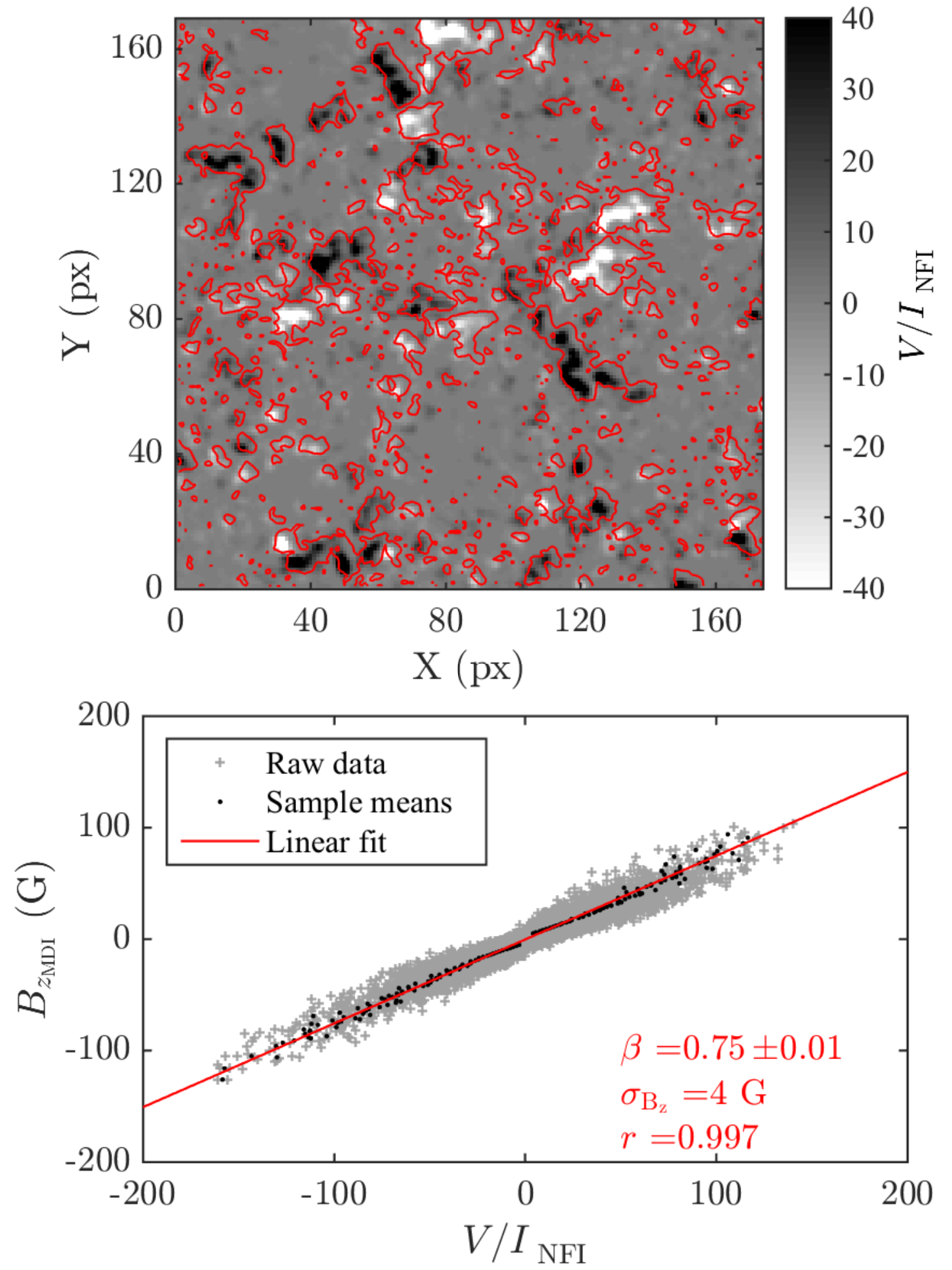}
	\caption{Top: NFI filtergram (right-hand side of the CCD) scaled between $-40\G$ and $+40\G$ with the red contours of the MDI magnetograms at $|\Bmdi| = 5\G$. The data are averaged over $4\hr$. Bottom: The light-gray crosses are the "$\px$ to $\px$" data. The dark dots are the binned data, fitted with a least-squares regression (red line). The fit parameters are defined as the slope $\beta$, the \mbox{$1\sigma$-uncertainty} $\sigma_{B_z}$ and the correlation coefficient $r$. }
\label{nfi_calib2}
\end{figure}

All the NFI data are resampled with the MDI pixel size of $0.6\arcs$. In addition, the resolution is degraded so that it is the same as MDI ($2\px = 1.2\arcs$), using a Gaussian convolution kernel with a FWHM of $1.2\arcs$. 
Next, in order to decrease the noise level in the MDI magnetograms ($\simm25\G$), we averaged both co-spatial data series over simultaneous time windows of $30\minutes$. This decreases the MDI noise level to $\simm5G$. Hence the NFI calibration only considers the pixels satisfying $|\Bmdi| \ge 5\G$. \Fig{nfi_calib2} (top) shows the areas that were finally used (i.e. the pixels in the magnetic patches within the red contours). In \Fig{nfi_calib2} (bottom) we have plotted $\Bmdi$ against $\VInfi$, pixel to pixel (gray dots). Note the spread of these data. Regardless of other instrumental effects (e.g. cross-talk and Doppler shifts in the line profiles), the spread of the scatter plot is mostly caused by the uncertainty of our co-alignment, which makes the NFI frames jitter around the MDI frames within a rather small, but non-negligible distance of $\simm0.6\arcs$ (i.e. displacements of $\pm 0.3\arcs$, see \S~\ref{sec:NFI_reg} or Table~\ref{table_coalignment}). Indeed, even a displacement of $1\px$ is enough to make a high flux density of a feature in one instrument correspond to a low flux in another instrument, this is particularly effective around the sharp edges of the magnetic features. However, we can reduce this spread by binning "vertically" the $\VInfi$ values, that is, averaging the $\VInfi$ values that fall within a $\Bmdi$ bin size of $1\G$. We obtain $N = 164$ independent pairs of data, plotted as black dots. Note that these points are much less spread out. They are fitted by the red line in \Fig{nfi_calib2} (bottom). The correlation coefficient is $r \approx 0.997$, the calibration coefficient equals $\beta = 0.75 \pm 0.01$, and the  \mbox{$1\sigma$-uncertainty} is $\sigma_{B_z} = 4\G$. Finally, we rescale the original NFI filtergrams and get calibrated "magnetograms" using $\Bnfi = \beta \times \VInfi$. We estimate the noise level in these calibrated, averaged magnetograms to be $\simm4\G$.
 	
\section{Flows, magnetic field, and X-ray emission \label{sec:multi_layer_analyses}}

Here we investigate in more detail the relationship between the photospheric flows, the magnetic field, and transient X-ray brightenings.
As mentioned, the photospheric flows were computed across the whole FOV from MDI continuum images. The flows across the SOT FOV were computed from the BFI blue continuum images.

\subsection{Detection of X-ray transients}
\begin{figure*}%/Users/attie/Matlab/raphael/codes/08Sep26/main_analysis/xrt_mdi_transient3_AAPaper2.m
	\centering
	\includegraphics[trim=0 0 0cm 0, clip, width = 1 \textwidth]{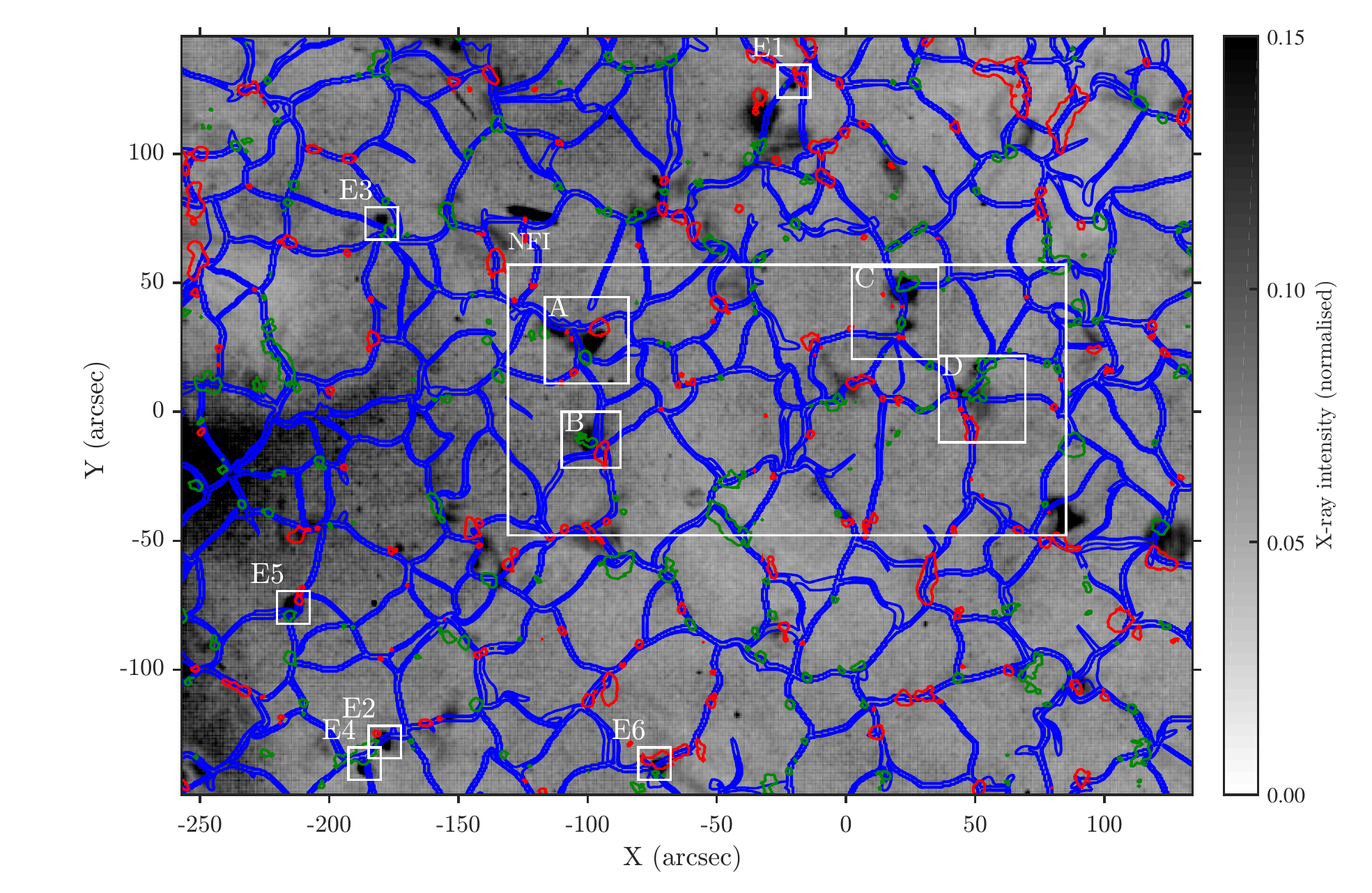}
	\caption{Mean of absolute running-difference XRT images, normalised by the average XRT image (\Eq{I_diff}). Each image in the series is taken $3\minutes$ apart from the previous and the next. The blue contours are the supergranular lanes, from the 4-hours averaged MDI flow field. The green and red contours (respectively) are the magnetic field strength at $-10\G$ / $+10\G$ from the $4\hr$-averaged MDI magnetogram. The white rectangle is the FOV of SOT (\Fig{overview3}). The smaller white rectangle encompass the X-ray transients. \label{xrt_mdi1}}
\end{figure*}

\Fig{xrt_mdi1} shows the supergranular lanes (blue contours) derived from the $4\hr$-averaged flow map of MDI. The lanes are derived with the automated cell recognition algorithm from \citet{Potts08}. They are displayed on top of the average of the absolute value of the running-difference XRT images, normalised by the $4\hr$-average image. Magnetic contours (red and green) from the averaged MDI magnetogram outline the "magnetic context" of the whole time series. The XRT averaged running-difference enhances all the transient events that occur over $4\hr$ of observations. The running difference uses a time interval of $3\minutes$ between the differenced images. This processed image, defined as $I_{\mathrm{diff}}$, formally derives as :
\begin{equation}
I_{\mathrm{diff}} = \displaystyle \frac{\sum_{i=1}^{N}|I_{i+\Delta n} - I_i|}{N\, \overline{I}},
\label{I_diff}
\end{equation}
where $\overline{I}$ is the $4\hr$-average XRT image, $I_i$ is the $i^{th}$ original image in a time series of $N = 240$ images, and $\Delta n$ is the number of frames between two subtracted images. Here, $\Delta n = 6$ with a time interval of $30\seconds$ between each frame. Normalising by the average image $\overline{I}$ has a "flat-fielding" effect, and enhances the contrast of the features even further. Thus $I_{\mathrm{diff}}$ 
%has no dimension and 
is expressed as a normalised intensity ratio.
The choice of the ideal time interval was made iteratively, by checking which interval reveals best all the short-lived emission, while smoothing out the long-lasting hot structures like X-ray loops, sigmoids, etc. X-ray emission clearly visible in \Fig{xrt_mdi1} comes from a source whose emission significantly increases over the background over $3\minutes$, and which we call  "transients".  In $I_{\mathrm{diff}}$, X-ray loops with variable emission may also still be visible. 

In the FOV of \Fig{xrt_mdi1} we looked for events whose intensity rises 15\% above their background level during less than an hour. We found 11 transients satisfying these criteria in the MDI FOV (small white rectangles, events A-D and E1 to E6), of which 6 are found in the NFI FOV, located in regions A-D, with multiple transients occurring in region C and D. Some small dots are also visible, and have all the characteristics of cosmic rays (1 pixel wide, present in 1 frame only with saturating intensity). The automated removal of the cosmic rays with the XRT software of Solarsoft is not possible as it also affects the data of interest. 
All the transients are located on the network, and are associated with barely resolved bipoles. In this respect they are quite similar to the ones studied by \citet{Krucker97}. Nonetheless, we cannot relate these events to those of \citet{Innes09}, as we do not have co-spatial and co-temporal observations to check any EUV counterpart to the observed X-ray transients.

We are able to obtain the thermal energies and underlying flows for all transients, but the MDI magnetograms were too poorly resolved and too noisy to quantify the signals accurately. Therefore in the next section, after first showing the sites of the MDI transients, we concentrate on those seen in the SOT FOV where the NFI magnetograms allowed an accurate tracking of the magnetic flux prior and during the transients using the magnetic balltracking technique \citep{Attie2015a}.

%\begin{landscape}
%\end{landscape}

\subsection{Transients in MDI-XRT FOV}

\begin{figure*}%/Users/attie/Matlab/raphael/codes/08Sep26/main_analysis/xrt_mdi_transient4_AAPaper2.m
	\centering
	\subfloat{%
	\includegraphics[width = \columnwidth]{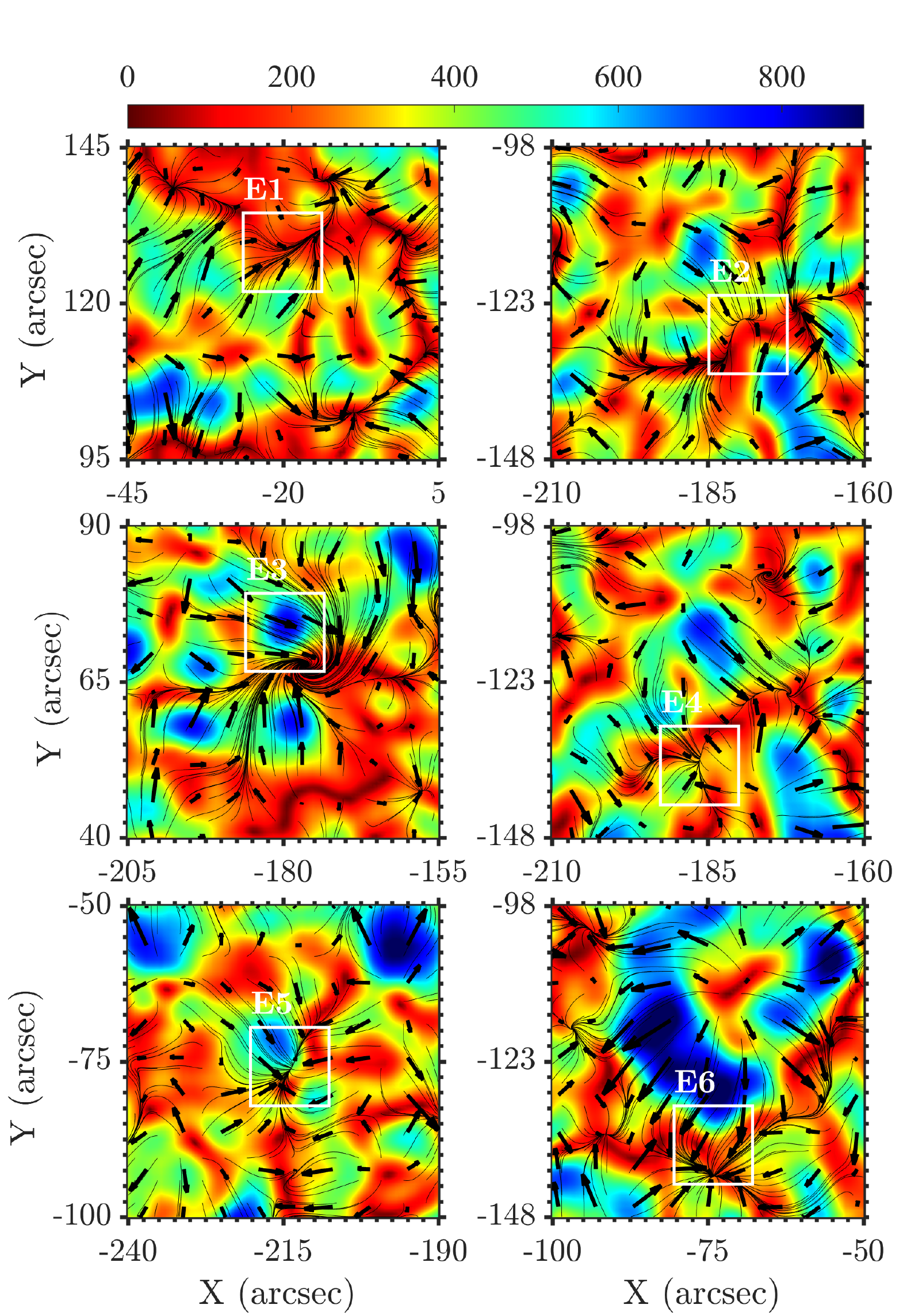}}
	\subfloat{%
	\includegraphics[width = \columnwidth]{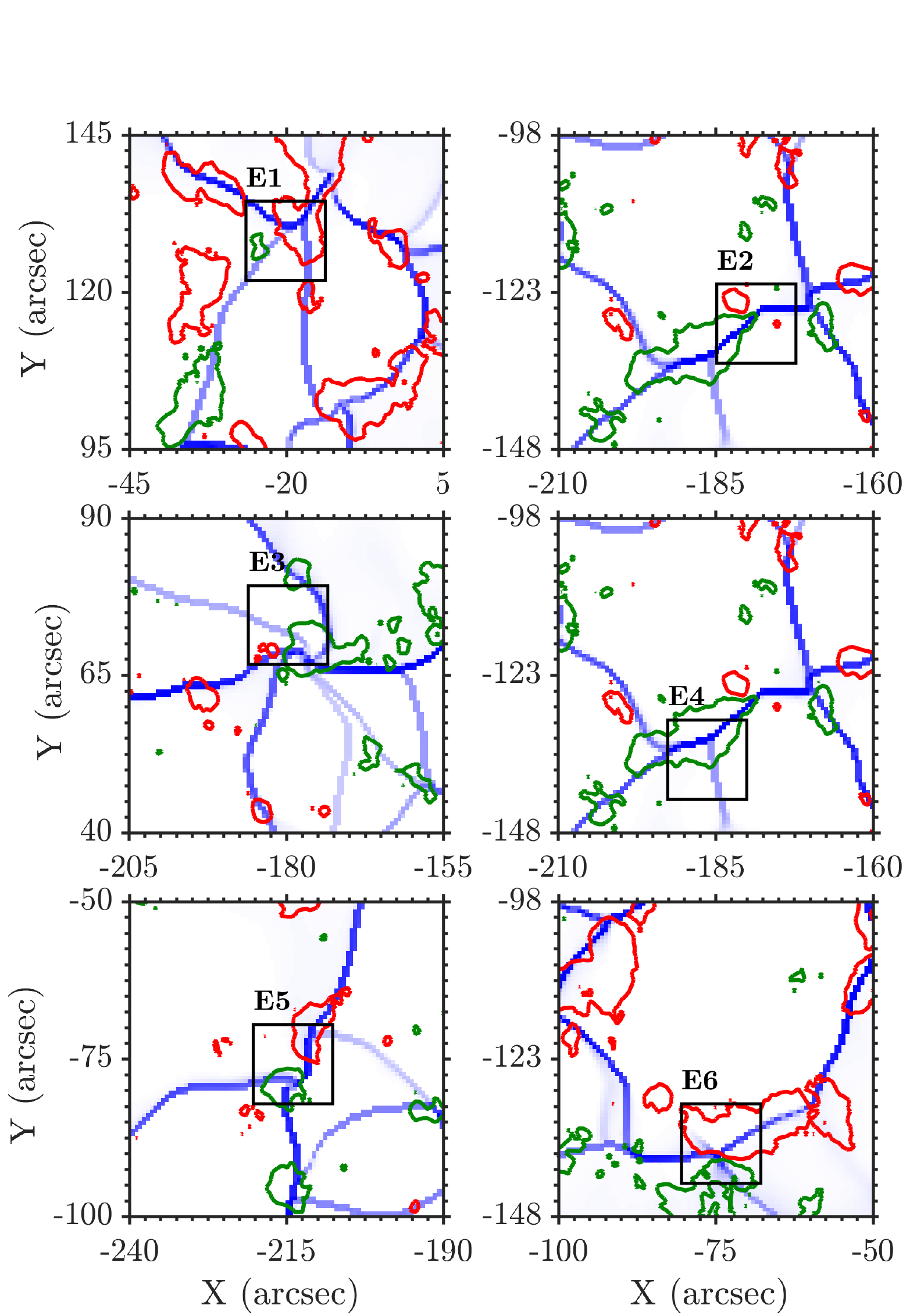}}
	\caption{Left: Close-ups on the flow at the sites of the events E1 till E6 shown in \Fig{xrt_mdi1}. The flows are averaged over $1\hr$ and smoothed over $4\Mm$. The coloured background is the magnitude of the horizontal flow velocity. The black lines are the streamlines and the arrows indicate the flow direction. Right: Same FOVs with the supergranular boundaries displayed as blue lanes; the green and red contours (respectively) are the magnetic field at $-5\G$ and $+5\G$ from the $4\hr$-averaged MDI magnetogram. The blue lanes are darker when they surround bigger, more persistent supergranules, and lighter for smaller, noisier supergranular boundaries.}
	 \label{Vtransients}
\end{figure*}

The flow fields are shown in \Fig{Vtransients} (left), where they are averaged over $1\hr$ and smoothed over $4\Mm$. The coloured background represents the magnitude of the horizontal velocity and the streamlines are drawn as black lines. Use \Fig{Vtransients} (right) as a complementary view of the supergranular boundaries (blue lanes), with the contours of the $4\hr$-averaged MDI magnetogram. We can see that the emitting sources E1 to E6 are not just located over the network (\Fig{xrt_mdi1}), but within groups of converging streamlines that are literally "funneling" the photospheric material right at the footpoints of the X-ray emission. The brightenings do not occur at random places within the network. Instead, they are located near the intersections of the supergranular network lanes (which we identify as the "crossroads" of several blue lanes in \Fig{xrt_mdi1}), with the exception of E2, which may be located in the middle of a network lane. 
The flow is particularly intriguing at the site E3 which is caught in one of the funnels of neighbouring supergranules whose streamlines get intertwined to form a supergranular vortex flow of about $25\arcs$ ($\simm18\Mm$).

\subsection{Transients in SOT-XRT FOV \label{SOT_transients}}

\begin{figure*}
	\centering
	\subfloat{%
	\includegraphics[width =\textwidth]{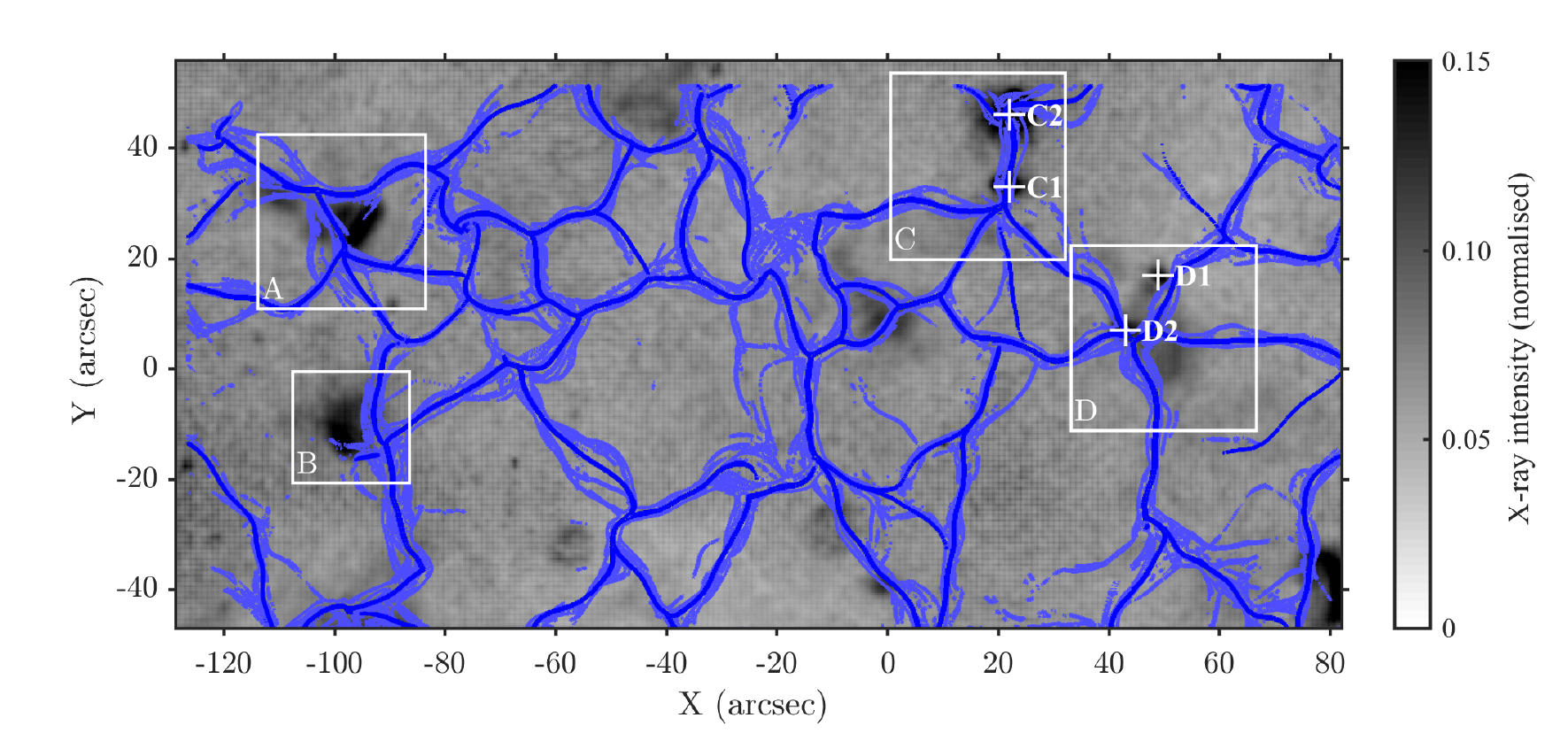}}
	
	\subfloat{%
	\includegraphics[width = \textwidth]{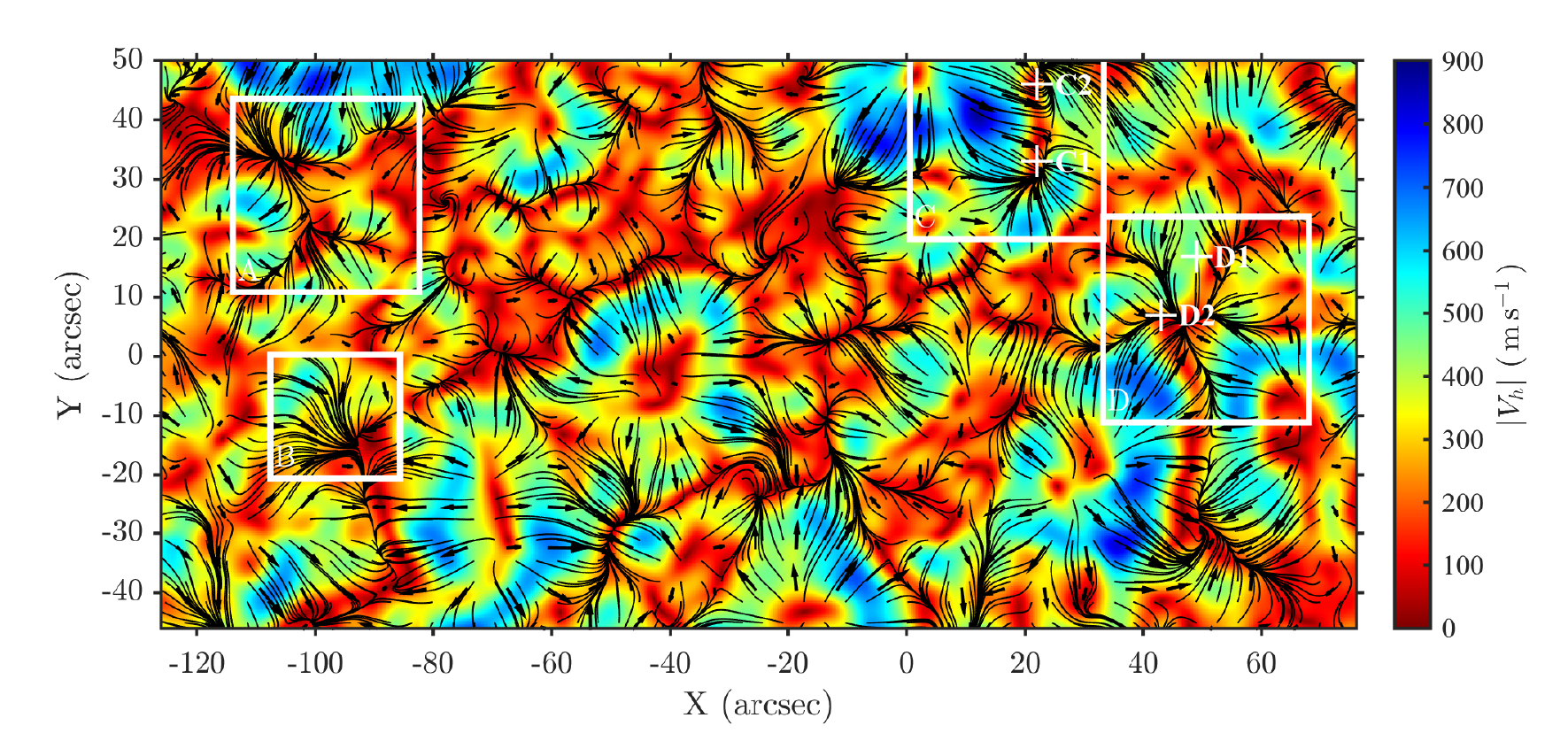}}
	\caption{Top: The blue lanes represent the supergranular network lanes. The gray background is the 4-hours-averaged absolute running-difference image from XRT. Bottom: Velocity field associated with the SOT field of view. The flow map is an average over 4-hours. The coloured background is the magnitude of the horizontal velocity. The small black arrows show the orientation of the flow.} \label{overview3}
	\end{figure*}

With the SOT data sets, the flow fields were smoothed over $4\Mm$ and averaged over $60\minutes$, and we derived the supergranular network lanes of each flow field. The different "snapshots" of the lanes were averaged over the whole time series, providing a context map of the flows and of the network, displayed in \Fig{overview3} (top). Region A and B contain somewhat elongated features and are probably barely resolved X-ray loops. In regions C and D we identified 4 sites of X-ray transients, tagged in white (C1, C2, D1, D2). The location of their emission peaks are tagged with white crosses. Note the preferred sites of the X-ray transients, with respect to the supergranular lanes: C1, C2, and D2 lie right on top of the intersection of the lanes. D1 is at the middle of a lane, that is, at mid-course between two intersections. A complementary view of the flow is given in \Fig{overview3} (bottom), averaged over $4\hr$. The transients are located in the funnelled streamlines like we observed previously in the MDI FOV. 

Below we describe in more detail the observations in regions C and D. Regions A and B are observed with the left half of the NFI CCD where, as mentioned in Section~\ref{sec:NFI_rejection}, calibration issues prevented us from quantitative measurements of the magnetic flux. The magnetic and X-ray evolution of all transients in the SOT FOV exhibited similar characteristics, so we only describe two: one in region C and one in D.

%/Users/attie/Matlab/raphael/codes/08Sep26/main_analysis/main_xrt_nfi_mballtrack_AAPaper2.m 

%/Users/attie/Matlab/raphael/codes/08Sep26/main_analysis/main_xrt_nfi_mballtrack_AAPaper2.m 
%\begin{figure*}
%\centering
%	\includegraphics[width = 1\textwidth]{./figures2/xrt_sot_velocity.pdf}
%	\caption{Velocity field associated with the SOT field of view. The flow map is an average over 4-hours. The colored background is the magnitude of the horizontal velocity. The small black arrows show the orientation of the flow. \label{overview4} }
%\end{figure*}

\subsubsection*{Region C and D}

In what follows, the magnetic balltracking combined with the region growing algorithm \citep{Attie2015a} was used to measure the flux evolution underneath the transients. In the following case studies, we can only discuss the evolution of positive flux. Indeed, the negative flux is spread out over too many magnetic elements that fragment and coalesce repeatedly, and they cover much larger areas than the positive flux. Local emergence and cancellation of negative flux, if any at all, is masked by a simultaneous decrease and increase of flux when multiple fragments are coalescing. For this reason, we can only discuss the evolution of the positive flux.

In region C, the magnetic threshold of the region growing algorithm is set to $10\G$. So a small percentage of flux may not be accounted for during the spatial integration. However, in region D, we could set it right above the noise level ($5\G$).

\begin{figure*}%/Users/attie/IDL/routines/main_scripts/08Sep26/cases/xrt_nfi_streamlines3.pro
	\centering
		\includegraphics[width = \textwidth]{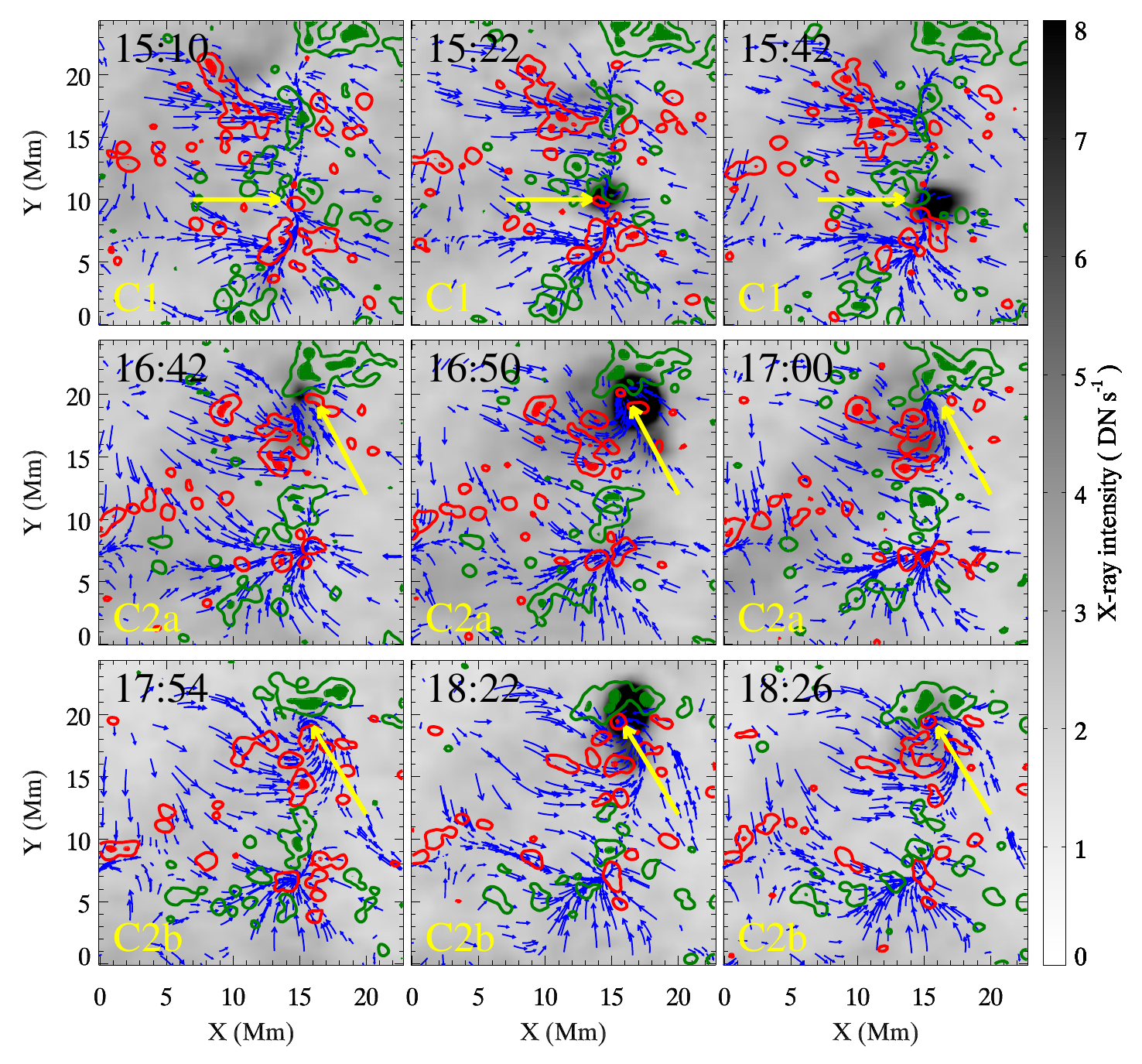}
	\caption{X-ray images (gray color table) and contours of the NFI magnetograms in region C of \Fig{overview3}. Red/green contours outline positive/negative polarity magnetic flux, respectively. Filled contours follow +/-$40\G$, thin contours +/-$10\G$. The blue arrows are the velocity vectors and show the direction of the flow. Their length is scaled linearly with the magnitude of the flow. The yellow arrows point at the location of the X-ray transients (C1, C2a, C2b).}
	\label{snapsC}
\end{figure*}

Several snapshots showing the three transients in region C (C1, C2a, C2b) are displayed in Fig.\ref{snapsC} (one event per row), with the blue arrows of the $60\minutes$-averaged flow fields, and the X-ray images in the background. These transients are observed between concentrations of positive and negative flux. 

\begin{figure*}%/Users/attie/IDL/routines/main_scripts/08Sep26/cases/main_xrt_nfi_mballtrack_AAPaper2.m
	\centering
	\includegraphics[width = 0.8 \textwidth]{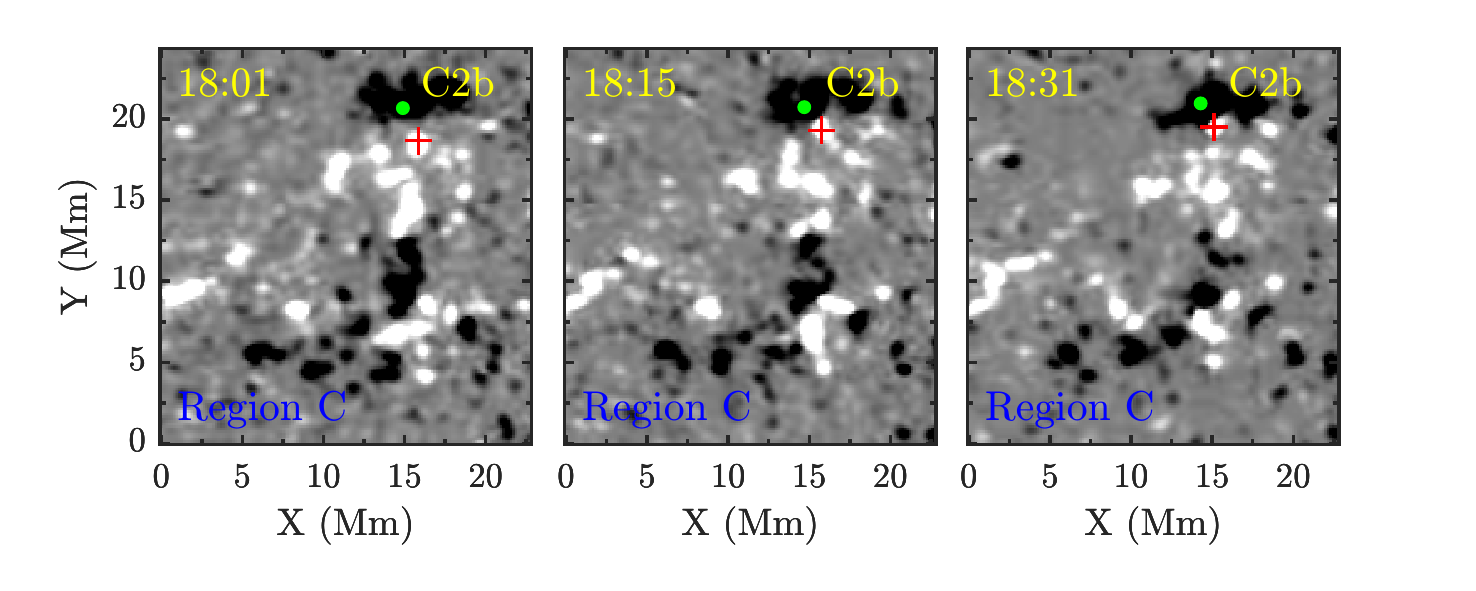}\\
	\includegraphics[width = 0.8 \textwidth]{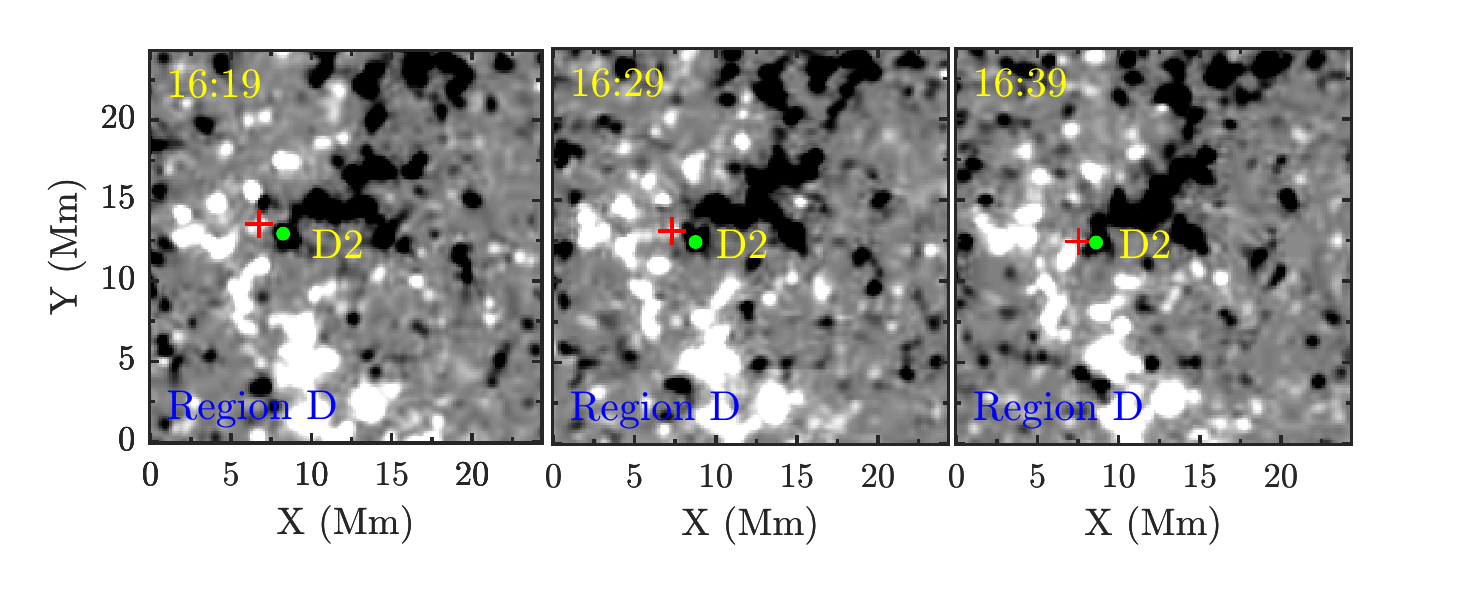}
	\caption{Magnetograms during the transients C2b (top) and D2 (bottom). The red crosses (resp. green dot)  are plotted at the centre of the ball tracking the fragment of positive flux (resp. negative flux).}
	\label{snapsCD}
\end{figure*}

\begin{figure} %/Users/attie/Matlab/raphael/codes/08Sep26/main_analysis/main_xrt_nfi_mballtrack_AAPaper2.m
	\centering
	\includegraphics[width = 1 \columnwidth]{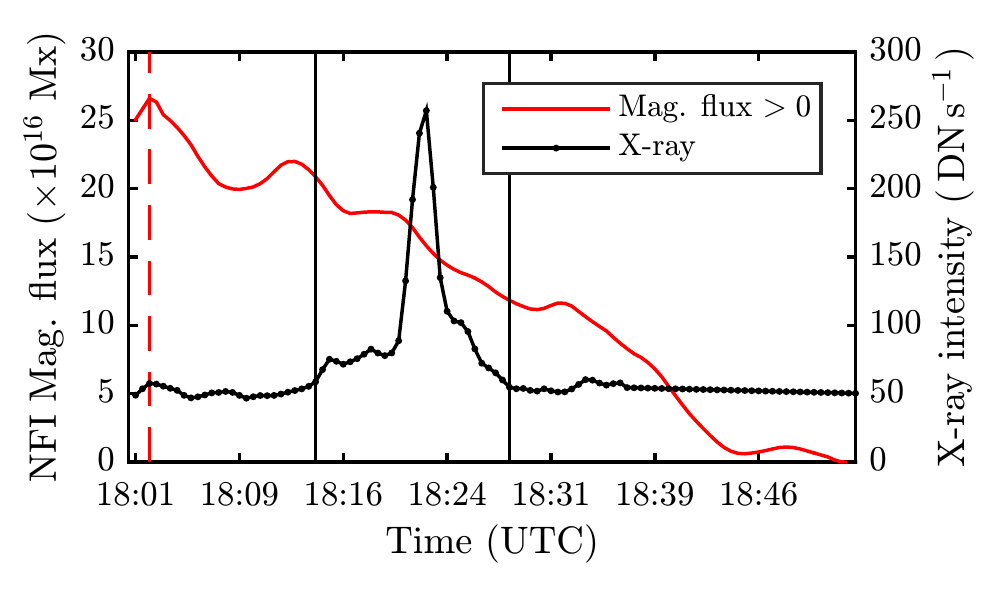}
	\caption{X-ray light curves with associated evolution of  positive magnetic flux for the transient C2b. The red dashed vertical line points at the local maximum of the flux. The black vertical lines show the beginning and the end of the transient. \label{fluxC2b}}
\end{figure}

\paragraph{Transient C2b}
The results of the magnetic balltracking for C2b are given in \Fig{fluxC2b}. The magnetic fragment that was tracked is visible in \Fig{snapsCD} (top). The light curve of the X-ray transient (black) is integrated spatially over the emitting source. The red curve is the positive magnetic flux derived from the magnetic balltracking. The red dashed vertical line points at the local maximum of the flux, and defines the beginning of the flux disappearance. The black vertical lines show the beginning and the end of the transient, defined as the time during which the light curve is 15\% greater than the averaged background emission. We can see that the X-ray intensity increases after the underlying magnetic positive flux has started to decrease. 

\begin{figure} %/Users/attie/Matlab/raphael/codes/08Sep26/main_analysis/make_velocity_figures.m
	\centering
	\includegraphics[width = 1 \columnwidth]{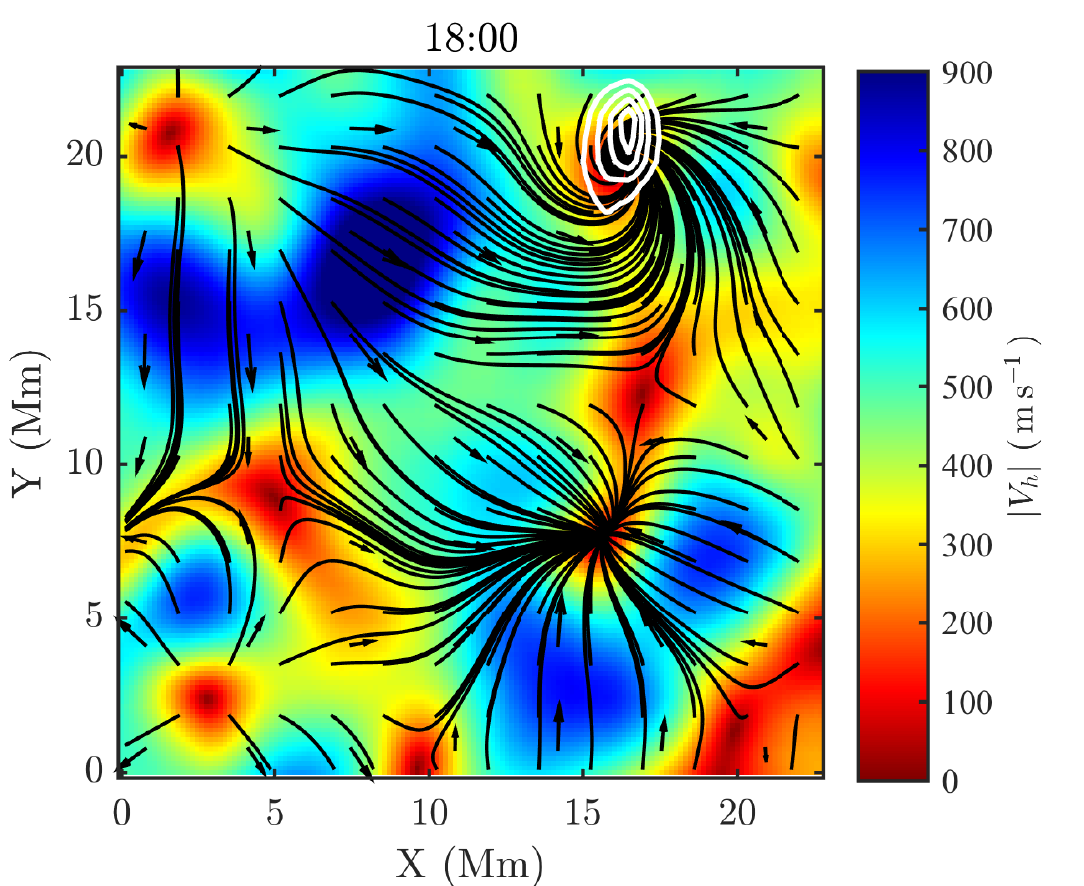}
	\caption{Flow field around 18:00 (C2b), averaged over $40\minutes$. The white contours are from the X-ray image at 18:22. Levels set between 10 and $25\DNS$. \label{velC_1800}}
\end{figure}

The positive magnetic flux is maximum at 18:02 with $\simm2.6\times10^{17}\Mx$. The X-ray transient starts at $\sim$18:14 while the magnetic flux has decreased by $\simm$20\%. The X-ray emission is maximum at 18:22 ($240\DNS$). 
\Fig{velC_1800} reveals that the X-ray transient occurs near the centre of a vortex flow formed at the northern intersection of the network lanes. The locations of C2a and C2b are close to each other, less than $2\Mm$ (which is below the $4\Mm$ resolution of the flow field), suggesting that this vortex was associated with multiple brightenings.

\begin{figure*} %/Users/attie/IDL/routines/main_scripts/08Sep26/cases/xrt_nfi_streamlines3.pro
	\centering
	\includegraphics[width = 1 \textwidth]{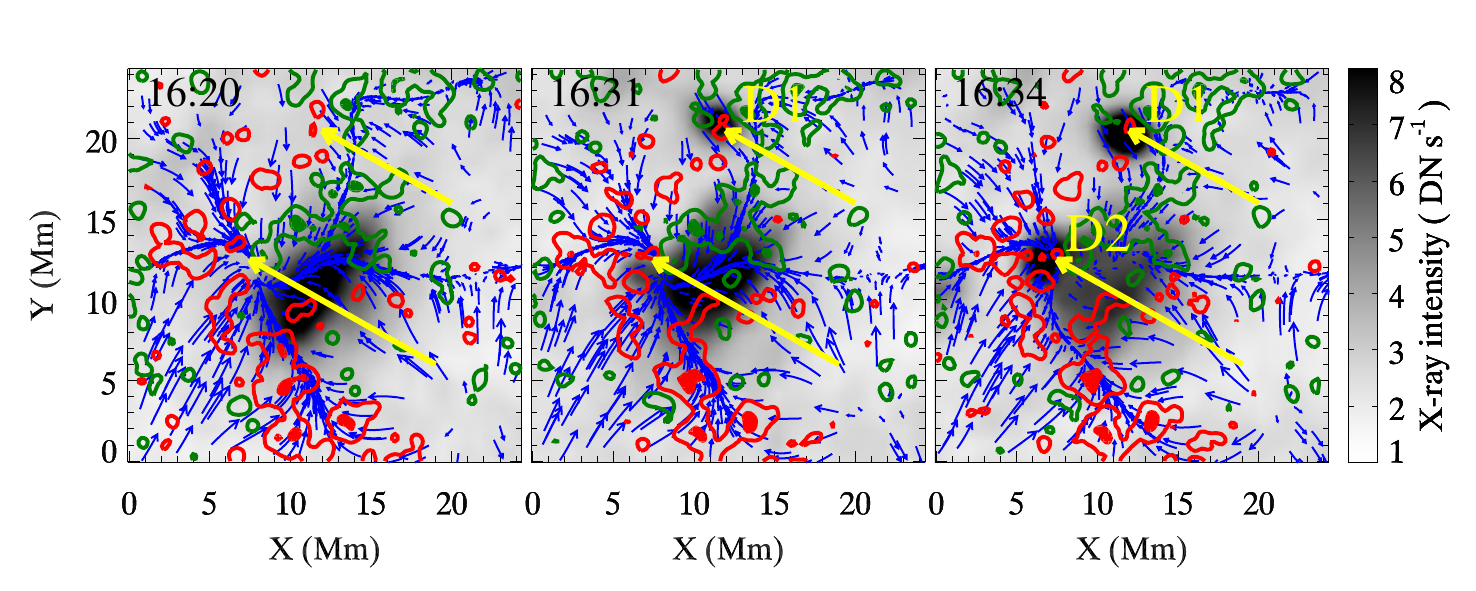}
	\caption{Snapshots of the X-ray time-series in region D. The yellow arrows point at the location of the X-ray transients. The top arrow points at the location of transient D1 (16:31), the bottom arrow to transient D2 (16:34). The blue arrows are the velocity vectors and show the direction of the flow. Their length is scaled linearly with the magnitude of the flow.} 
	\label{snapsD}
\end{figure*}

\paragraph{Transient D2} Most of the X-ray emission in region D comes from barely resolved X-ray loops near the centre of the snapshots plotted in \Fig{snapsD}. However the transient D2 is located near the footpoints of these loops. The transient D2 is best visible in the third snapshot of \Fig{snapsD}. 

\begin{figure}%/Users/attie/Matlab/raphael/codes/08Sep26/main_analysis/main_xrt_nfi_mballtrack.m
	\centering
	\includegraphics[width = 1 \columnwidth]{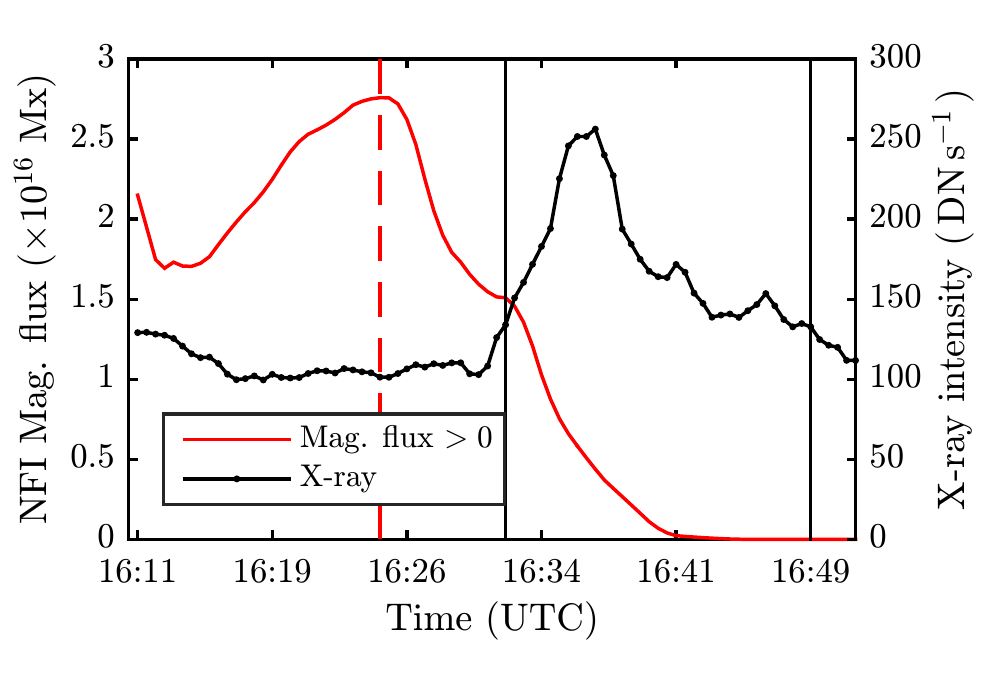}
	\caption{X-ray light curve of the transient D2, with the associated underlying flux cancellation. The red dashed vertical line points at the local maximum of the flux. The black vertical lines show the beginning and the end of the transient.}
	\label{fluxD}
\end{figure}

The magnetic fragments that were tracked underneath are visible in \Fig{snapsCD} (bottom). The threshold of the region growing algorithm is set right above the noise level, $\simm5\G$. Even at this threshold, the magnetic balltracking was still able to track and extract the disappearing magnetic feature from the surrounding. \Fig{fluxD} shows that the magnetic flux is maximum at 16:25 with \mbox{$\simm3.4\e{16}\Mx$}. It has decreased by $\simm$25\% at 16:31, when the X-ray transient begins. The X-ray emission is maximum at 16:36 ($265\DNS$).  

The secondary maximum of the magnetic flux in transient D2 at 16:31 comes from oscillations in the whole FOV, and are not specific to these magnetic features. The oscillations are more visible here because the flux density of the tracked features is weaker, on average, and integrated over smaller areas than in the previous cases.

\begin{figure} %/Users/attie/Matlab/raphael/codes/08Sep26/main_analysis/main_xrt_nfi_mballtrack.m
\centering
		\includegraphics[width = 1 \columnwidth]{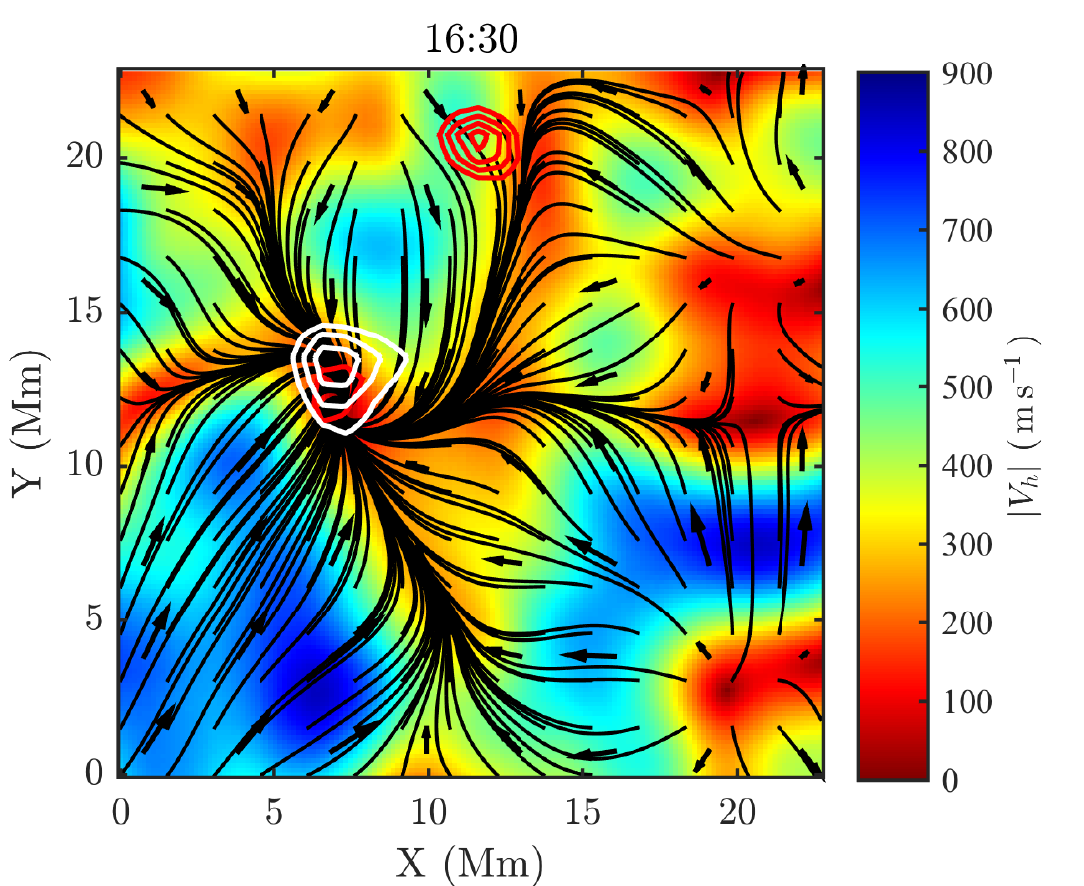}
	\caption{Flow field in region D averaged over $40\minutes$ centred on 16:30. The red contours are from the X-ray image (Fig.\ref{snapsD}) at 16:31, white contours at 16:34. Contour levels are set between $10$ and $20\DNS$. \label{velD_1630}}
\end{figure}

D2 is right at the intersection of supergranular lanes according to \Figs{overview3} (top) and \ref{velD_1630}. The streamlines seem to twist as they converge. This vortical structure is caused by the unbalanced velocity from opposite sides of the supergranule boundaries. The velocity is on average greater than $550\mets$ within $5\Mm$ from the supergranular lanes of the lower left supergranule, while it is slower (less than $450\mets$) in the other supergranules. 

All five X-ray transients appeared shortly after the magnetic flux starts to "disappear", and were situated right between opposite polarities. Thus we believe that the X-ray emission signifies magnetic reconnection in the low corona. As this occurs at the supergranular boundaries, that is, regions of downflows, we cannot rule out that the observed magnetic cancellation is, in fact, due to the submergence of these loops, regardless of any reconnection process higher up. 
% Thus we believe that they signify magnetic cancellation resulting from magnetic reconnection. 

% that is made more likely by the converging flows, and the close encounters of loops that can create a current sheet, in the same manner as the so-called Moving Magnetic Features, MMFs, in \citet{Shimizu02}

\subsection{Energy of the X-ray transients}
\begin{figure}%/Users/attie/IDL/routines/hinode/XRT/response_function2.pro
\centering
	\includegraphics[trim=1cm 0cm 0cm 0cm, width= 1 \columnwidth]{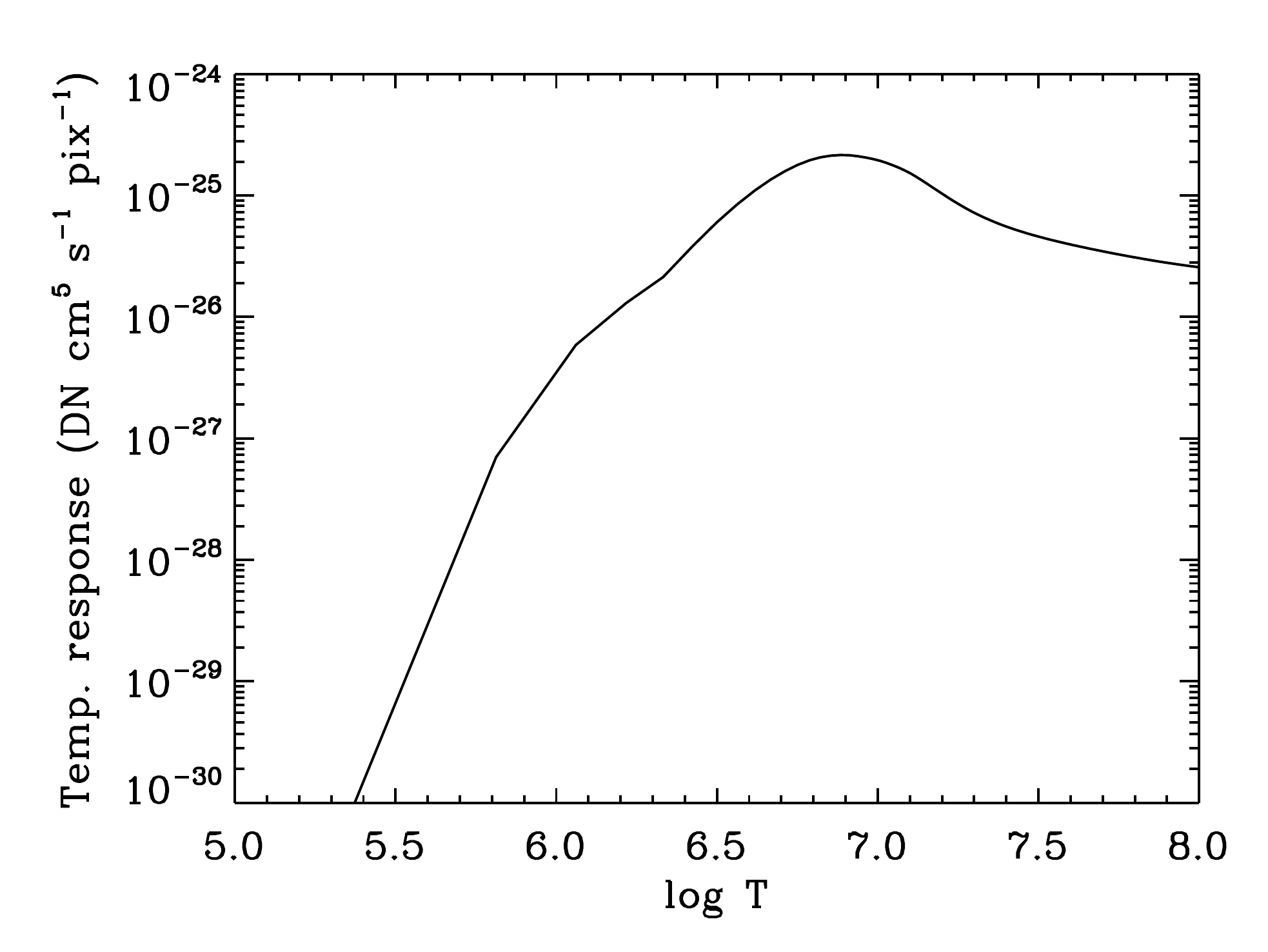}
	\caption{Temperature response of XRT with the C-poly filter.  \label{chianti}}
\end{figure}

Quiet-Sun soft X-ray sources in the magnetic network have already been reported by \citet{Krucker97}, and called "network flares". Because of their observational similarities (time and spatial scales), we followed the same method to calculate the energy released by these X-ray sources. We assume a temperature $T=1.2\,\mathrm{MK}$, and integrate a synthetic coronal spectrum using the CHIANTI package \citep{CHIANTI1997,CHIANTI2009}, with the XRT response functions corresponding to the C-poly filter, provided by the XRT software in Solarsoft. 
The emission measure is calculated using the relation 
\begin{equation}
I_{obs}^{C_{poly}}(T) \sim EM(T)\int_{\nu(C_{poly})} J_\nu(\nu,T) ~ \epsilon(\nu) d\nu,
\label{EM1}
\end{equation}
where $I_{obs}^{C_{poly}}(T)$ is the observed intensity at a given temperature $T$, $EM(T)$ is the emission measure, $J_\nu$ the synthetic spectrum from CHIANTI, calculated with the procedure "isothermal.pro", and $\epsilon$ the spectral response of XRT associated with the C-poly filter. 
The integral on the right hand side of \Eq{EM1} is the "temperature response", and is shown in \Fig{chianti}.

$EM(T)$ is defined along the line-of-sight, and is proportional to the squared density of the electrons ${n_e}^2$  times the source length $d_z$:
\begin{equation}
EM(T) \approx {n_e}^2 d_z \quad \mathrm{(cm^{-5})}.
\label{EM2}
\end{equation}

The electron density is therefore $n_e = \sqrt{EM(T) / d_z}$ ($\mathrm{cm^{-3}}$).  If we assume that particles have been heated from chromospheric temperatures to provide the X-ray emission, and are filling a cubic volume of side length $d_z=d$, then the total number of particles in the volume would be:
\begin{equation}
N = n_e d^3 = \sqrt{EM(T) d^5} 
\label{Nparticles}
\end{equation}

Hence the thermal energy of an X-ray emitting source at temperature $T$ on the solar surface, as used for the network flares in \citet{Krucker97},
\begin{equation}
E_{th}=\frac{3}{2}N k T \approx \frac{3}{2} k T \sqrt{EM(T)\,d^5}.
\label{energy}
\end{equation}

\begin{figure}%/Users/attie/IDL/routines/hinode/XRT/response_function2.pro
\centering
	\includegraphics[trim=1cm 0cm 0cm 0cm, width= 1 \columnwidth]{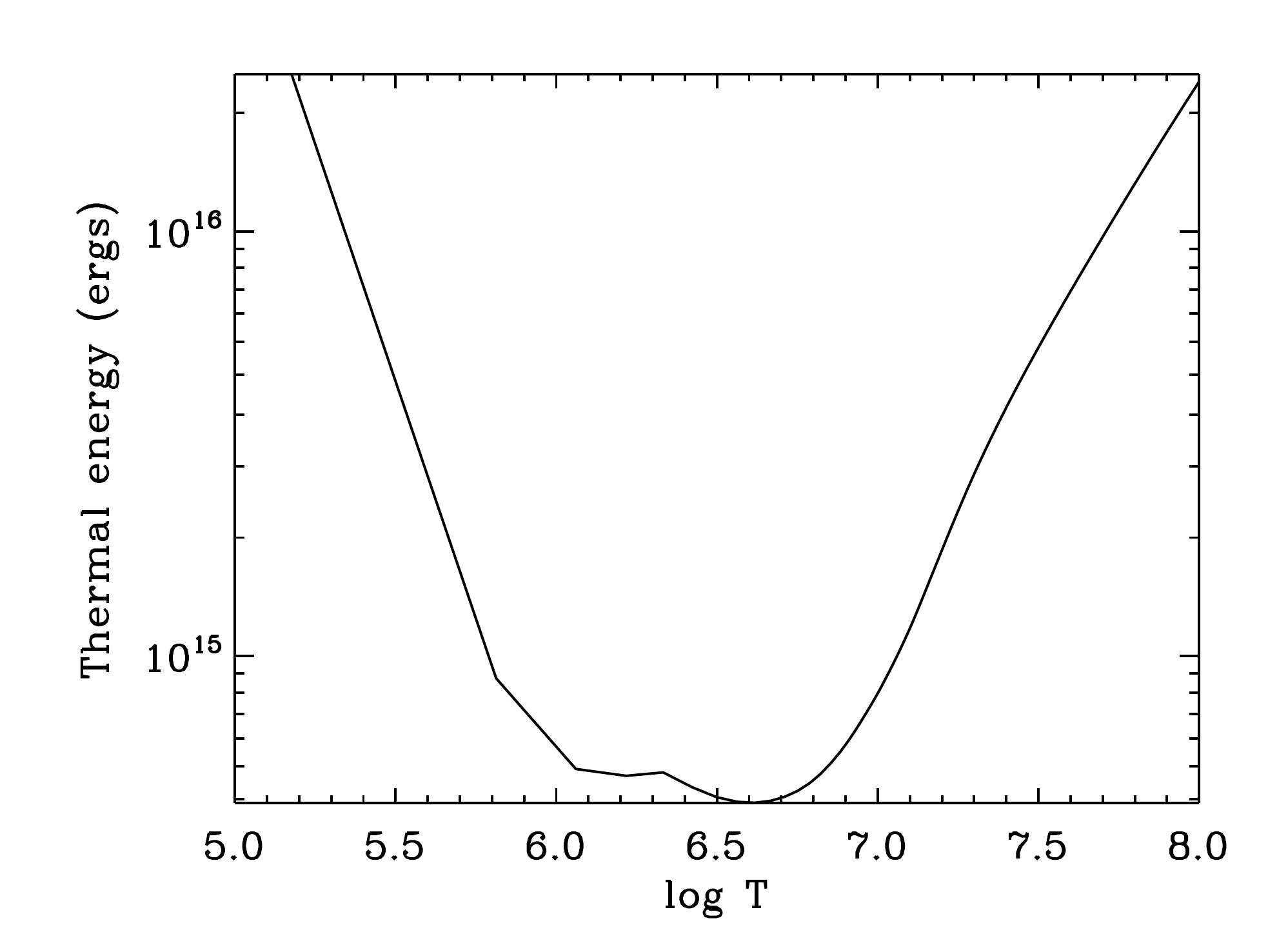}
	\caption{Thermal energy as a function of the temperature for $I_{obs}^{C_{poly}}(T)=1\,\mathrm{DN}$ and \mbox{$d=0.15\,\mathrm{Mm}$}. \label{energy_dependence}}
\end{figure}

We only have observations through one filter. So it is not possible to obtain the temperature using filter ratios, and we take the same temperature of $1.2\MK$ as in \citet{Krucker97}. Nonetheless, the energy defined in \Eq{energy} has a very weak dependence on the temperature between $1\MK$ and $10\MK$. This is shown in \Fig{energy_dependence} where we used $I_{obs}^{C_{poly}}(T)=1\,\mathrm{DN}$ to compute the energy dependence versus the temperature. For this, we used a source size of $d=0.15\,\mathrm{Mm}$ (i.e. the pixel size of our resampled images).

Over a large temperature range of $1$ to $10\MK$, the energy varies by 50\%. It seems very unlikely that the temperature(s) of the source exceeds this temperature range. The uncertainty in the thermal energy depends much more on the shape and the size of the source, by a factor of $d^{2.5}$ according to equation \ref{energy}, rather than on the choice of the temperature, provided it stays in the range $1$-$10\MK$. While the horizontal extension of the source of several pixels is a known parameter, its dimension ($d_z$) along the line-of-sight  is unknown. From this we expect an uncertainty of at least one order of magnitude on the energy calculated thereafter.

To compute the energy from the XRT images, the intensity at each pixel is inserted in \Eq{EM1} to obtain the emission measure $EM(T)$. The emission measure is spatially integrated over a square with side length $d$. The latter is measured as the averaged Full-Width-at-Half-Maximum of the 2D emitting structure at its maximum emission.
The background emission measure is retrieved by averaging $EM(T)$ over the time preceding the flaring phase of the transients. We remind that the flaring phase was previously defined as when the X-ray intensity is 15\% above the background. $EM(T)$ is then integrated over the time of the flaring phase, and the background is subtracted.
We finally obtain a net increase in emission measure $\Delta EM(T)$ which represents the amount of material heated to the temperature $T$, which is inserted in \Eq{energy}. 
This process was repeated for each transient, in the MDI FOV and SOT FOV. 

\subsubsection*{Results}

\begin{table}%/Users/attie/Matlab/raphael/codes/08Sep26/main_analysis/xrt_mdi_transient2.m
\caption{Thermal energy released from sources E1 till E6 assuming $T=1.2\,\mathrm{MK}$. \label{table1}}
	\begin{center}
		\begin{tabular}{ccccccc}
		\hline
		Transients & E1 & E2 & E3 & E4 & E5 & E6\\	
		\hline	
		$E_{th}$($10^{25}\,\mathrm{ergs}$) & 20.4 & 38.3 & 23.7 & 6.7 & 16.3 & 30.3\\
		$d$ ($\Mm$) & 1.1 & 1.3 & 1.7 & 1.0 & 1.1 & 1.2\\
		\hline
		\end{tabular}		
	\end{center}
\end{table}

\begin{table*}%/Users/attie/Matlab/raphael/codes/08Sep26/main_analysis/main_xrt_nfi_mballtrack.m
\caption{Parameters associated with the X-ray transients assuming $T=1.2\MK$.  \label{table2}}
\begin{center}	
	\begin{tabular}{l c c c c c}	
	\hline
	Transients & $E_{th}$ ($10^{25}\erg$)  & $d$ ($\Mm$)  & $\Delta \phi_p/\phi_p$  (\%) & $t_{reac}$ ($\mathrm{min}$) & $\Delta t_{tr}$ ($\mathrm{min}$) \\
	\hline
	C1	& 1.3 & 0.4 & 30 & 3 & 29 \\
	C2a  & 1.8 & 0.4 & 10 & 4.5 & 14 \\
	C2b  & 1.1 & 0.4 & 20 & 12 & 14 \\
	D1    & 1.9 & 0.5 & 35 &  8  & 12 \\
	D2    & 2.6 & 0.7 & 40 & 6.5 & 17 \\
	\hline
	\end{tabular}
\end{center}
\tablefoot{
$E_{th}$: thermal energy; $d$: source size; $\Delta \phi_p/phi_p$: percentage of positive flux cancellation; $t_{reac}$: reaction time between the beginning of the X-ray transient and the beginning of the magnetic cancellation; $\Delta t_{tr}$: transient lifetime.
}	
\end{table*}	

The results are summarised in Tables \ref{table1} and \ref{table2}. The energies of the network flares in the NFI FOV are on average smaller ($10^{25}\erg$, Table~\ref{table2}) by one order of magnitude than the ones outside the NFI FOV ($10^{26}\erg$, Table~\ref{table1}). In each case there is an uncertainty of one order of magnitude due to the longitudinal source size $d_z$ (in the direction of the line-of-sight) that is unknown. The smaller energy in the NFI FOV is mainly due to the smaller dimensions of the sources (at least, on the horizontal dimensions). 
%This difference from one instrument to the other is explained by the fact that Hinode is given a specific "target-region" when planning the observations. In the present case, we asked for "quiet Sun". Therefore, the Hinode science plan induced a selection effect as the pointing was on a region with smaller, less energetic sources than MDI. The latter has a wider FOV and includes relatively more energetic events. 

Below, we calculate the average energy flux released by the transients in the NFI FOV, which are the least energetic, and the energy flux released outside the NFI FOV, whose X-ray sources are more intense by about one order of magnitude.
\begin{itemize}
\item The NFI FOV covers an area of $6\times10^3\Mm^2$ of the quiet Sun, during $4\hr$ of observations. The total energy flux averaged over this area and this time duration, released by the 5 transients C1,C2a, C2b, D1, and D2 is of the order of $10\,\mathrm{erg\,s^{-1}\,{cm}^{-2}}$.
\item The MDI FOV (minus the area covered by the NFI FOV) covers an area of $3.5\times10^{4}\Mm^2$. Which gives an averaged energy flux released by the 6 transients E1 to E6 of the order of $10^2\,\mathrm{erg\,s^{-1}\,{cm}^{-2}}$. 
\end{itemize}
We only selected transient events whose emission suddenly increases by 15\% above the background emission within minutes. They do not represent all the possible sources of X-ray emission. Nevertheless, based on the 11 transients studied here, we have estimated that the average energy flux is between $10$ and $10^2\,\mathrm{erg\,s^{-1}\,{cm}^{-2}}$ in the quiet Sun. %  FOV = 1.4.10^10 cm wide, 2e20 cm2
This is $10^2$ to $10^3$ times less than the minimum coronal heating requirement of the quiet Sun (between $10^4$ and $10^5\,\mathrm{erg\,s^{-1}\,{cm}^{-2}}$) \citep{WN77,Aschwanden2004}.

\section{Discussion}

\begin{figure}%inkscape
	\centering
	\includegraphics[width=1 \columnwidth]{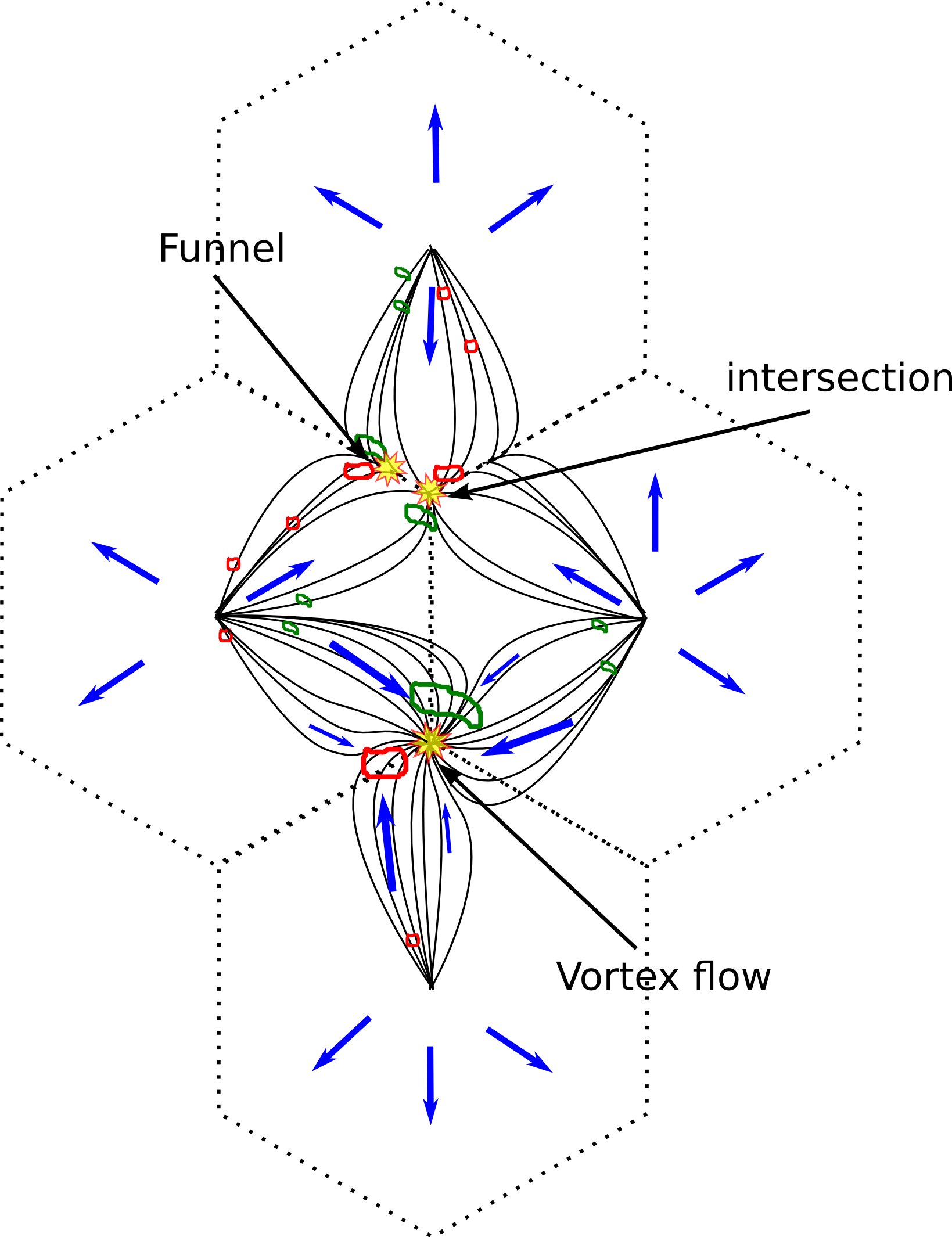}
	\caption{Sketch of the different steps leading to an X-ray transient. The hexagonal dashed lines represent idealised supergranular boundaries. The black lines are streamlines of the flow. The blue arrows show the main orientation of the flow. Bigger arrows symbolise faster flows than the smaller arrows. The yellow stars represent X-ray transient events. The green and red thick contours represent magnetic features of opposite polarity.}
	\label{cartoon2}
\end{figure}

In the previous sections we have described 11 X-ray network flares, and the photospheric flows underneath. With SOT/NFI, we could also quantify the cancellation of the magnetic flux which was followed by the flaring of the X-ray transients. For the 6 other transients, due to the lower resolution and lower signal-to-noise ratio of the MDI magnetograms, the evolution of the bipolar magnetic fields at the flaring sites could not be reliably followed. Yet we could still define similar patterns in the flows that seem to drive the magnetic flux towards a seemingly inevitable cancellation. These patterns are sketched in the cartoon of Fig.\ref{cartoon2} that summarises in a single simplified picture the processes observed for the 11 X-ray transients, which are all associated with magnetic cancellation.

In Fig.\ref{cartoon2}, we represent the flaring sites at the places where the streamlines of the flow converge. These are  the intersections of the supergranular boundaries. We interpret the observed topology of the flows as follows: As the flows converge to either side of the network lanes, the streamlines form a funnel. Eventually, the supergranular flow becomes unbalanced, and the velocity in one side of a supergranular lane is greater than in its neighbouring supergranule. Consequently, antagonistic flows become asymmetric with respect to ideal boundaries. The streamlines are reshaped accordingly, and the direction of the resulting flow is determined by the average flow, with streamlines of the "dominant" supergranules pushing back the "weaker" ones. As long as the weaker flow does not accelerate, the directions are kept, and structures like large-scale vortex flows persist \citep{Brandt88,Attie09,Bonet2010}. Otherwise, they get disrupted and are barely visible in long-time-average flow fields. When the flow is more balanced, streamlines converge symmetrically to the intersection without noticeable vortical topology, forming funnels leading into the intersection.

In the next section, we discuss the possible effects of the observed topology of the flow on the dynamics of the magnetic flux. 

\subsection{Converging flux model}

The mechanism by which the flux decreases can be explained by the loop submergence below the photosphere and/or by the reconfiguration of the magnetic field  \citep{Kubo2007}; yet we can discuss that the occurence of the network flares may involve magnetic reconnection higher up  in a manner that falls within the converging flux model described in \citet{Priest94}. While this model describes the triggering mechanism of X-ray bright points (BPs) at larger scales than the present events, our observations are \textit{a priori} similar: two magnetic fragments of opposite polarities approach each other, cancel out, while an intense X-ray emission is observed.
The model explains the energy release as a result of the interaction of the magnetic fragments with the background field, which eventually leads to the formation of a current sheet and magnetic reconnection in the higher layers. 
In what follows, we use the magnetic balltracking to measure the parameters defined in this model, and to calculate the estimated energy released during the reconnection. We compare it with the energy $E_{th}$ released during the eruption of X-ray network flares (Table \ref{table2}). 

\subsubsection{Definition of the model parameters \label{sec:converging_model_params}}
For each event in the NFI FOV, the key parameters of the model are calculated with magnetic balltracking. They are illustrated in \Fig{CFMParameters}, and define as:

\begin{figure} %Drawn with Keynote
	\centering
	\includegraphics[width = 1 \columnwidth]{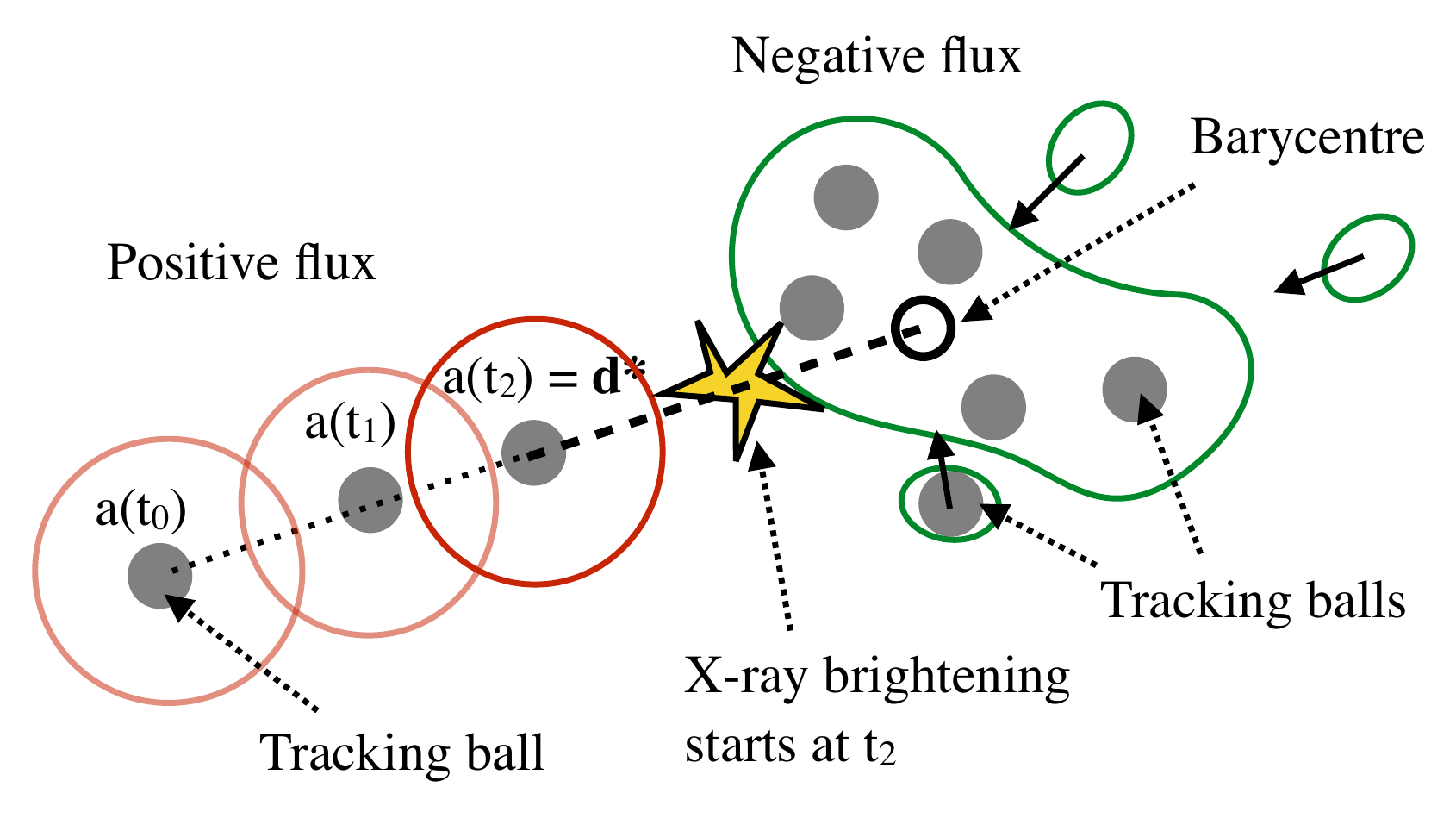}\\
	\includegraphics[width = 1 \columnwidth]{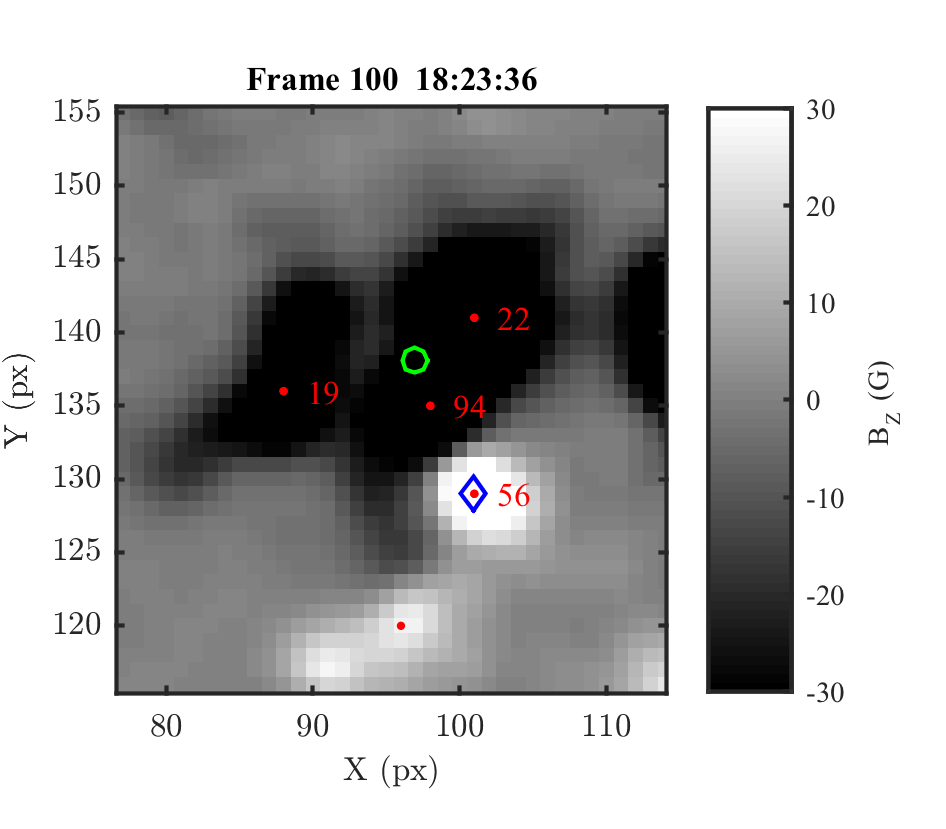}
	\caption{Top: Illustration of the Converging Flux Model parameters, as used in combination with the results of magnetic balltracking. The X-ray brightening starts at time $t_2$. This time is used to define the interaction distance $d^{\ast}$. The barycentre is the weighted average of the position of the tracking balls used in the algorithm. Bottom: Example with case C2b of the weighted average of the tracked position. The green circle is the barycentre of the 3 balls labeled 19, 22, 94. The blue diamond is the barycentre of ball 56; the only one tracking the white patch so it is at the same position as the ball centre (red dot).} 
	\label{CFMParameters}
\end{figure}

\begin{itemize}
\item The time-dependent distance $2\,a(t)$ between the centre of the two moving fragments of opposite polarities. 
 \item The approach speed $\text{-}\dot{a}$ of the fragment, which is half the time derivative of the above quantity.  
Calculating this requires the tracking of the positions of both fragments. As mentioned in \Section{SOT_transients}, for each of the transients in the NFI FOV, only the positive flux can be discussed here. Nonetheless, our algorithm can still track  the different local minima within the closest, and either clustered out or wider negative patches using a few balls. To do this we use their "barycentric" position. Here this so-called "barycentre" is the weighted average of the positions of the tracking balls, where the weight of each ball is equal to the absolute flux density at each ball position. This "barycentric tracking" is useful in the case of large magnetic features where more than one local minimum are found. This is illustrated in \Fig{CFMParameters} (bottom). This method reduces the ambiguity and the uncertainty in choosing which fragment (with negative flux) to use for calculating the interaction distance. Then we use this weighted-average position to derive the relative approach speed $\dot{a}$ of the fragment of positive flux with respect to the fragments of negative flux.
 \item The interaction distance\footnotemark $\,d^{\ast}$, defined in \citet{Priest94} as the distance from the middle of the two features centres to the point where the magnetic null point is formed, and projected onto the photosphere. It originally defines as:
 \begin{equation}
 d^{\ast} = \sqrt{\frac{f}{\pi B_0}},
 \label{interaction_distance}
 \end{equation}
where $f$ is the flux of the fragment, and $B_0$ is the intensity of the horizontal background field above the photosphere. In the model, the parameters are defined assuming symmetry, that is, with the unsigned flux of both fragment being strictly equal. This is not true in our observations and causes an error of more than one order of magnitude. Instead, we can directly use the results of magnetic balltracking to measure $d^{\ast}$. Our definition of the interaction distance $d^{\ast}$ is illustrated in \Fig{CFMParameters} (top) as the distance $a(t_2)$, and formerly defined as the half-length between the tracked barycentric positions of the fragments, at the time when the flux is at its maximum. This occurs a few minutes before the X-ray transient starts. 
\footnotetext{To avoid confusion with the X-ray source size $d$ used earlier in the paper, we note the interaction distance $d^{\ast}$ instead, also used in \citet{Priest94}.}

\item The magnetic fragment width $w$ measured with the region-growing algorithm by taking the average diameter of the extracted area of the fragment.

\item The cancellation time $\tau_c$, which is the time it takes for the flux of the fragments to completely cancel. It is defined as:
\begin{equation}
\tau_c = \frac{w}{\dot{a}}.
\end{equation}
\end{itemize}

With the above quantities, the free energy stored in the current sheet $\Wfree$ in excess of a potential field is defined as:
\begin{equation}
\Wfree=\frac{{B_0^2} {d^{\ast}}^3}{2\mu}\,F_s(a/{d^{\ast}}),
\label{Wfree}
\end{equation}
where $\mu$ is the permeability, and $F_s(a/{d^{\ast}})$ is a scaling factor determined numerically that depends on the ratio of $a$ and $d^{\ast}$ \citep[Eq. 3.28]{Priest94}. The scaling factor varies rapidly with $a/{d^{\ast}}$. It is equal, respectively, to 0.6, 2.5, and 4.4 when $a/{d^{\ast}}$ = 0.5, 0.2, and 0.1 ($a/d^{\ast}$ decreases when the fragments approach each other). We use the value of $a/d^{\ast}$ at the start time of the reconnection, which is here assumed to be at the beginning of the X-ray network flares.
In what follows, a horizontal background field $B_0$ of $5\G$ is used. This value is consistent with what we get from potential field extrapolation right above the photosphere in the regions of interest.

\begin{table*}%/Users/attie/Matlab/raphael/codes/08Sep26/main_analysis/main_xrt_nfi_mballtrack3_AAPaper2.m
\caption{Parameters related to the converging flux model in \citet{Priest94} and to the energy flux in \citet{Galsgaard1996}. \label{table3}}
\begin{center}	
	\begin{tabular}{l c c c c c c}	
	\hline
	Transients & $\Wfree$ ($\erg$) & $E_{\mathrm{F_p}}$ ($\erg$)  & $d^{\ast}$ ($\mathrm{Mm}$) & $\dot{a}$ ($\mathrm{m.s^{-1}}) $  & $w$ ($\mathrm{Mm}$) & $\tau_c $($\minutes$) \\
	\hline
	C1		&  $\esim{23}$ & $\esim{24}$ & 0.9 & 500 & 1.5  & 32 \\
	C2a		&  $\esim{23}$ & $\esim{24}$ & 1.1 & 700 & 1.6 & 24 \\
	C2b		&  $\esim{23}$ & $\esim{25}$ & 1.3 & 600 & 1.6 &  51 \\
	D1		&  $\esim{22}$ & $\esim{23}$ & 1.1 & 500 & 1.6 & 57 \\
	D2		&  $\esim{21}$ & $\esim{23}$ & 0.7 & 300 &  0.9 & 28 \\
	\hline
	\end{tabular}
\end{center}
\tablefoot{
$\Wfree$: "free" magnetic energy in excess of a potential field configuration; , $E_{\mathrm{F_p}}$: Energy released within sheared flows; $d^{\ast}$: interaction distance;  $\dot{a}$: absolute value of the approach speed; $w$: average fragment width; $\tau_c$: cancellation time.
}	
\end{table*}

\subsubsection{Results}

\begin{figure}%/Users/attie/Matlab/raphael/codes/08Sep26/main_analysis/main_xrt_nfi_mballtrack3_AAPaper2.m
	\centering
		\includegraphics[width = 1 \columnwidth]{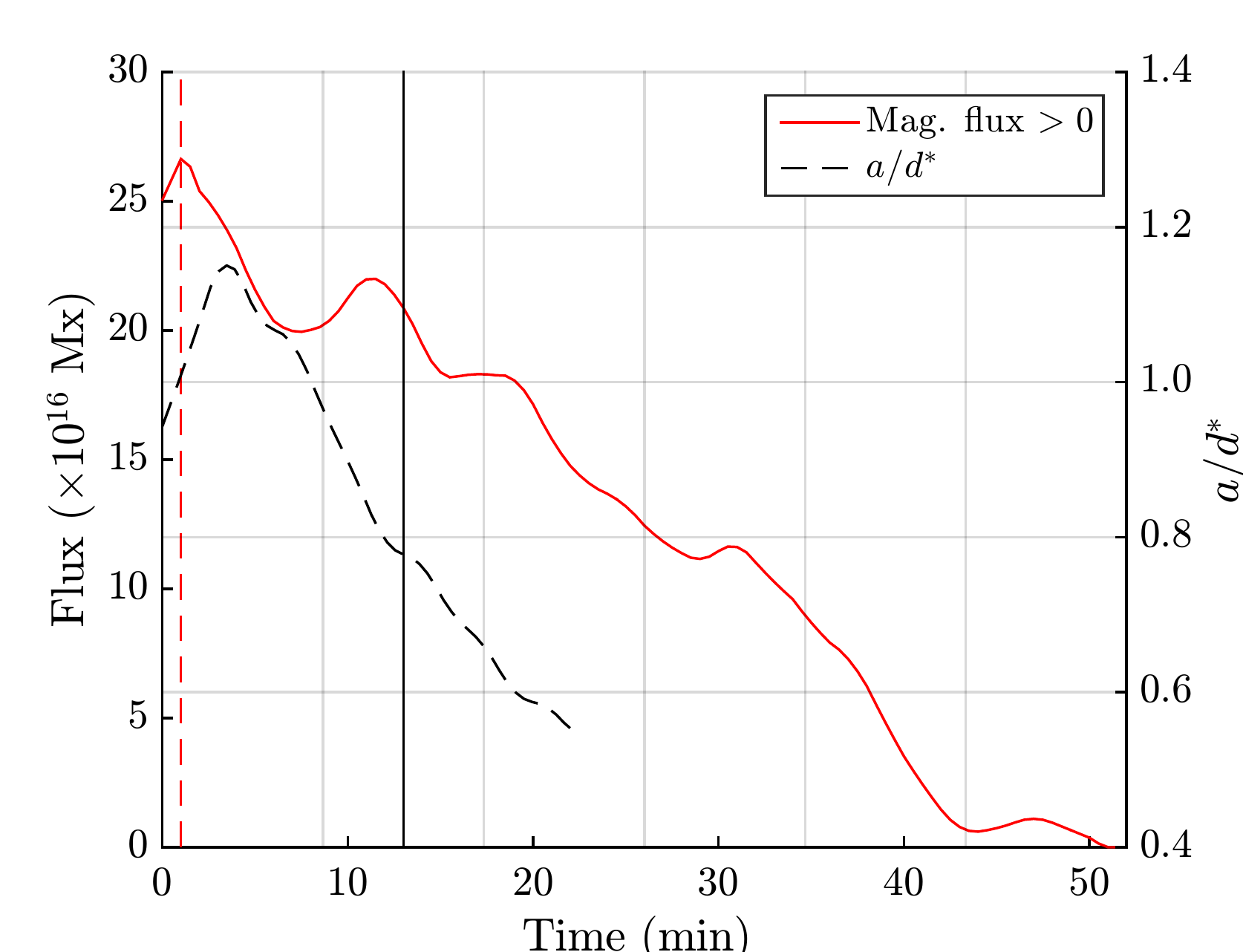}
	\caption{Evolution of the positive flux with $a/d^{\ast}$ for the transients C2b. The vertical discontinuous red line is at the maximum of the flux, and $a/d^{\ast}=1$. The black vertical line marks the beginning of the X-ray transient.}
	\label{distances_C}
\end{figure}

\begin{figure}%/Users/attie/Matlab/raphael/codes/08Sep26/main_analysis/main_xrt_nfi_mballtrack3_AAPaper2.m
	\centering
		\includegraphics[width = 1 \columnwidth]{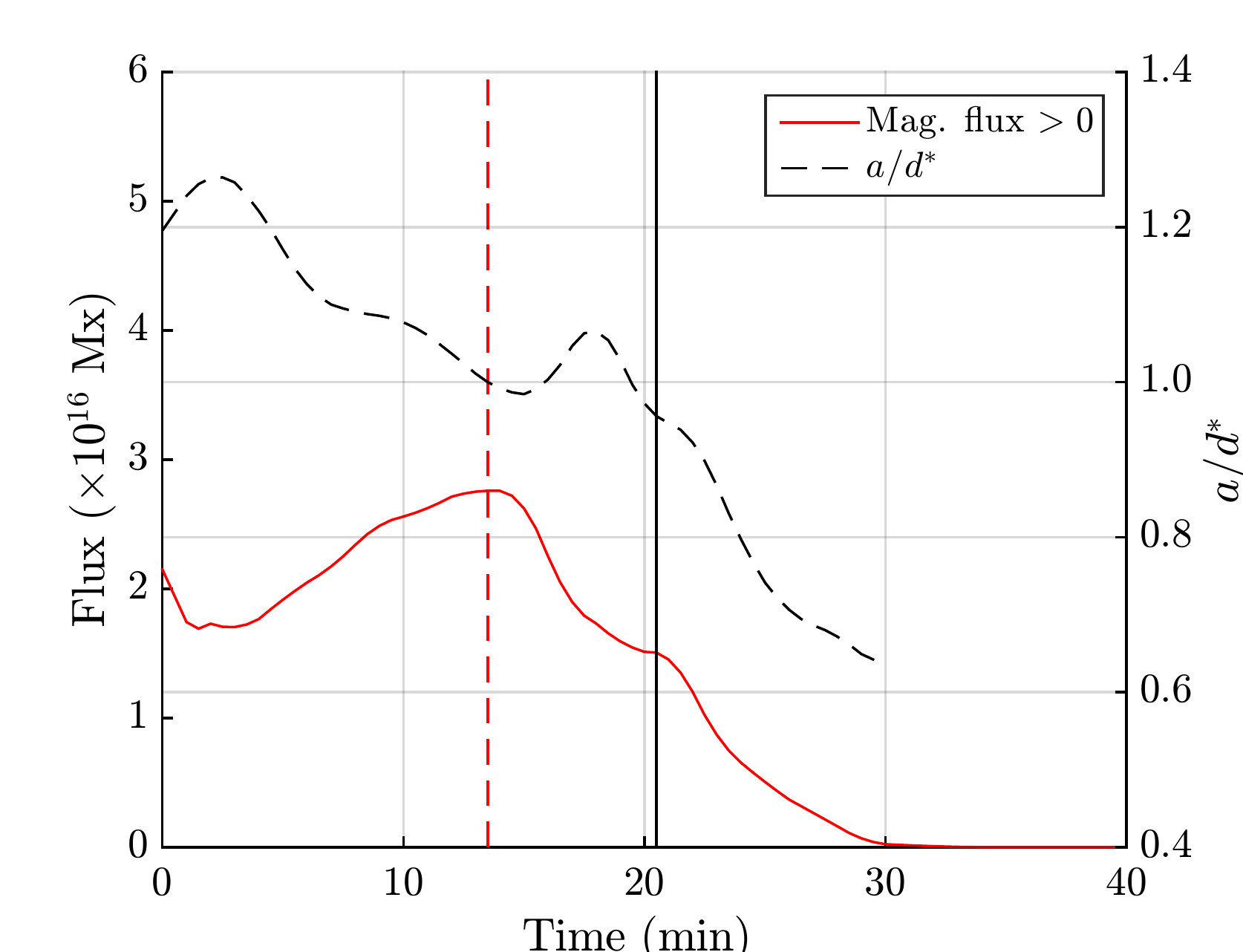}
	\caption{Same as Fig.\ref{distances_C} for the transient D2 in region D.}
	 \label{distances_D}
\end{figure}

The values of the above quantities are summarised in Table~\ref{table3}. The interaction distance $d^{\ast}$ varies from $0.7\Mm$ to $1.3\Mm$. \Figs{distances_C} and \ref{distances_D} represent the positive magnetic flux cancellation (red continuous line) calculated in \Section{SOT_transients}, along with the distance ratio $a/d^{\ast}$ (black discontinuous line) which is used in $F_s(a/{d^{\ast}})$ to scale the free energy $\Wfree$. The red discontinuous vertical line shows where the flux cancellation starts, which by definition corresponds to $a/d^{\ast} = 1$. The black vertical line marks the beginning of the X-ray transient and sets where we take the value of $a/{d^{\ast}}$ to compute the scaling factor $F_s$. The ratio $a/{d^{\ast}}$ is $\simm1.0$ at the start of the transients C2b, and D2. It scales $\Wfree$ by $F_s \sim  5\e{-2}, 3\e{-3}$ (respectively for C2b and D2). Finally we obtain $\Wfree$ of the order (resp.) $10^{23}$ and $10^{21} \erg$. So the free energy in the transient C2b, and D2 is significantly scaled down. This is due to the fact that the X-ray transient starts too soon after the fragment has moved past the interaction distance (i.e. $a/{d^{\ast}}$ is close to one). We observe, however, that the flux of the fragment beneath the transient is unbalanced, with a ratio of negative flux over positive flux equal to 10, and up to 80 in D2. So this is far from the symmetric topology assumed in the model (which assumes a symmetry with respect to the vertical axis). In \citet{Priest94} the opposite vertical magnetic field lines of the bipoles reconnect with each other, and the reconfiguration occurs within the background horizontal field ($B_0$), while the lower loop sinks into the photosphere. Here $B_0$ is quite low ($5\G$) which results in rather small free energy (\Eq{Wfree}), compared to the thermal energy.

\subsubsection{Comparisons of $E_{th}$ and $\Wfree$}
\begin{figure} %/Users/attie/Matlab/raphael/codes/08Sep26/main_analysis/main_xrt_nfi_mballtrack3.m
	\centering
		\includegraphics[width = 1 \columnwidth]{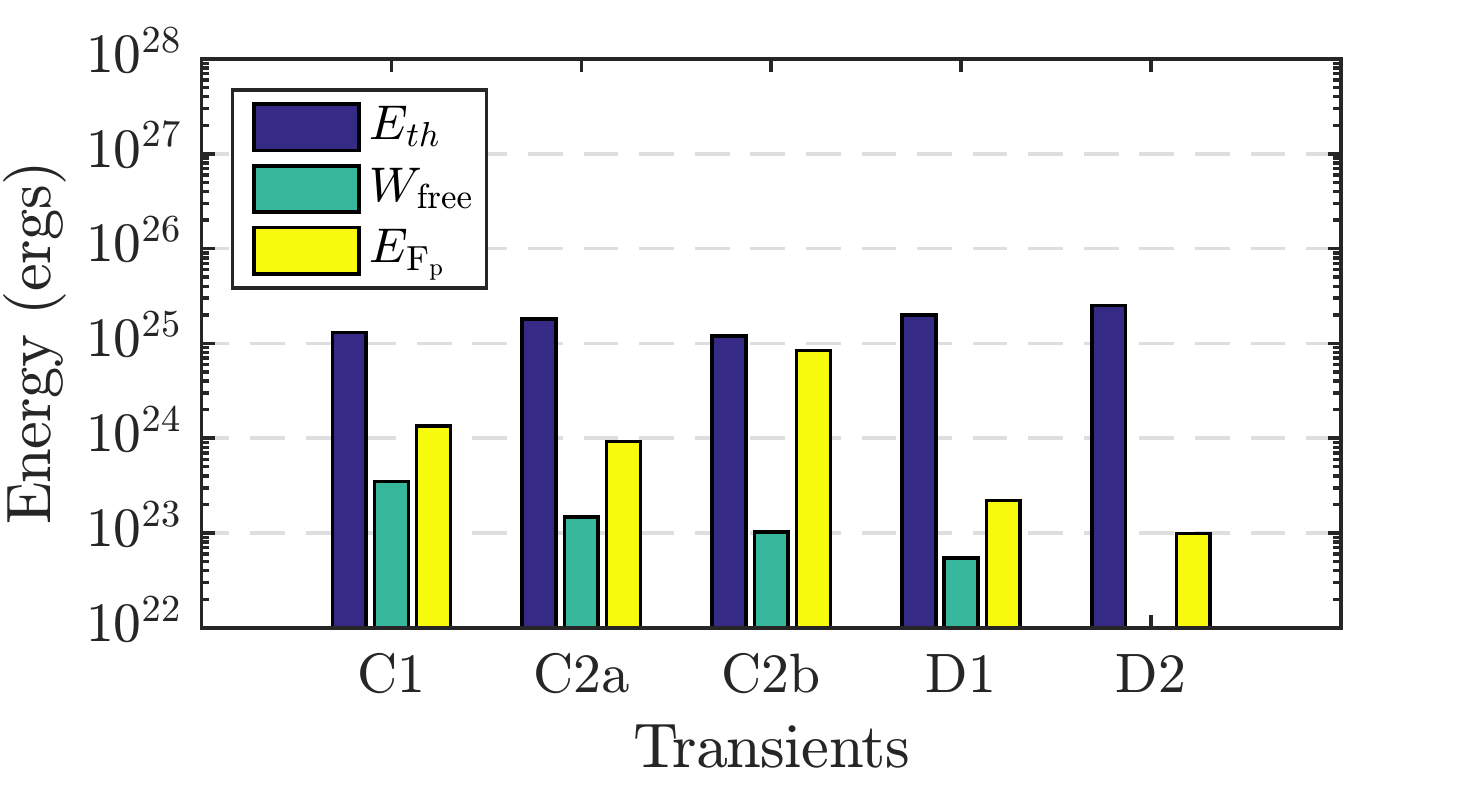}
	\caption{Bar plot of the thermal energy $E_{th}$, the free energy $W_{free}$ of the converging flux model, and the average Poynting flux $\langle F_p \rangle$ for the transients in region C and D.}
	\label{energy_comparison}
\end{figure}

The thermal energy $E_{th}$ measured from the X-ray emission and the free magnetic energy $\Wfree$ calculated with the flux convergence model are compared in the bar plot in \Fig{energy_comparison}. The third energy $\langle F_p \rangle$ will be discussed in the next section. As we mentioned in the previous paragraph, the main source of errors in the calculation of the free energy lies in the actual field topology (imbalance between the positive flux and negative flux)  which might be quite different from the ideal case (symmetrical). This affects the estimation of $\Wfree$ as a function of $B_0^2$. The non-linear dependence of the scaling factor $F_s$ on $a/{d^{\ast}}$ is another source of uncertainty when tracking multiple local minima which affects the calculation for C2a and C2b: the difference between the position of the geometric centre and the weighted centre of the tracked local minima (\Fig{CFMParameters}) impairs $F_s$ by up to one order of magnitude. We also remind that there is an uncertainty of one order of magnitude when calculating $E_{th}$ (due to the source size). Nonetheless, on average, the thermal energies are all greater than the free energy. This is particularly clear with D1 and D2 where the free energy is negligible compared to the thermal energy. 
Therefore, even if within 1 order of magnitude, the piecewise potential field configuration invoked in the converging flux model provides sufficient energy for a later release during the transients C1, C2a, C2b, it is more unlikely to do so in D1 and D2. Therefore one must investigate other possible sources of energy. 

\subsection{Effects of the funnels and the vortices \label{sec:shears}}

\subsubsection{Shearing motions}

\citet{Galsgaard1996} have studied the effect of the shearing of an initial homogenous magnetic field. It was shown that the longer systematic shearing acts on the field, the greater the free energy. This is caused by the exponential growth of currents caused by the field lines bending and converging to a more confined area. 
\\In our observations, we have emphasised the presence of supergranular vortices and twisted funnels. In fact, these are the sites of higher shear than in the relatively more laminar flow of the internetwork. In such configurations, we can consider the time and spatial scales of the funnels and the vortices observed at the erupting sites in region C and D, and apply them to the model of average energy dissipation per unit area and per unit time in \citet{Galsgaard1996}, roughly equal to the average Poynting flux $\langle F_p \rangle$. With $B_z$ the intensity of the vertical magnetic field, $V_d$ the velocity advecting the field lines, and $\phi$ the inclination angle of the field lines, in \mbox{centimetre-gram-second (cgs) units} we have:
\begin{equation}
\langle F_p \rangle = \frac{B_z^2\, V_d \, \tan(\phi)}{4 \pi}.
\label{Fp}
\end{equation}
Here, we use $B_z = <\Bnfi>$, that is, the mean vertical magnetic field of the fragment with cancelling flux. We consider the fragments tracked in region C and D. We also define $t_d$ as the characteristic time of the shear motions acting on the field lines, and $L$ the characteristic size of the region over which the Pointing flux is integrated. If we take $L= 5\Mm$, which is the characteristic length of the funnels along which the magnetic fragments are transported, $\tan(\phi)$ is approximated by:
\begin{equation}
	\tan(\phi) \sim \phi \sim V_d t_d / L .
\end{equation}

The above quantities were already calculated in \S~\ref{sec:converging_model_params} (Table~\ref{table3}). In fact, we use $t_d = \tau_c$ as a lower limit, where $\tau_c$ was defined as the cancellation time, that is, the time it takes for the flux to vanish. $V_d$ is set to the approach speed $\dot{a}$. We note that without the actual geometry of the 3D magnetic field for an accurate measurement of $\tan(\phi)$, the latter is also a source of uncertainty, on which $\langle F_p \rangle$ depends linearly.

Finally, we integrate the energy flux $\langle F_p \rangle$ over the region of sheared flow (of size L) and over the cancellation time $\tau_c$ to get an order of magnitude estimate of the total dissipated energy $E_{\mathrm{F_p}}$ from the magnetic fragments caught in the vortices and the funnels. $E_{\mathrm{F_p}}$ is represented by the third bar plot in \Fig{energy_comparison}. The energy appears greater than in the converging flux model. However, if we account for the uncertainty in determining the horizontal component of the field in the converging flux model, there is no clear difference between $\Wfree$ and $E_{\mathrm{F_p}}$ in the cases C1 and C2a. Nonetheless, the fact that \Eq{Fp} accounts for the actual observed flux (and not the extrapolated horizontal component) leaves much less uncertainty than in the converging flux model, so that for C2b, D1 and D2 the dissipated energy $E_{\mathrm{F_p}}$ is greater than $\Wfree$.  

\citet{Galsgaard1996} emphasised that persistent shearing may be the source of a bursty regime of the energy release, and we believe that this indeed is the case for transient C2a and C2b which occur near the centre of a vortex flow (see Figs.\ref{snapsC},\ref{velC_1800}). 

\subsubsection{Large-scale vortices and funnels}

Due to the presence of large-scale vortices and twisted funnels transporting the magnetic flux to the junctions of the lanes, we can also comment on the results of the simulations from \citet{Amari00,Amari2003a,Amari2010} with non-zero-helicity magnetic field, that are specific implementations of the more general Flux Cancellation Model from \citet{vanBallegooijen1989}. Although it is applied to coronal mass ejections, the initial states used in the model are in many aspects the same as observed here. In these simulations, the amount of total flux that cancels is within a broad range of 6\% to 30\%. This broad range is believed to be caused by the different amount of shear given to the magnetic field at the initial state. In the real situation, vortices and funnels are shearing the magnetic field, and a broad range of cancelled magnetic flux could be expected as well. In the 5 X-ray transients that we analysed (C1, C2a, C2b, D1, D2), the amount of cancelled flux ranges between 10\% to 40\% (see $\Delta \phi_p/\phi_p$  in Table~\ref{table2}). This is only the longitudinal flux, and we do not make any assumption on the amount of cancellation of the transverse component of the flux. 
In the simulations, the flux cancellation is due to small-scale mixing and reconnection that is followed by the formation of a flux rope, which gets disrupted in the end. While we have no direct observations of the flux rope, the presence of funnels and the vortices at the pre-interactive phase of the converging flux model are conditions quite similar to the simulations, where the same topology of the flows is used. 

The energy release is believed to occur through Joule dissipation. The kinetic energy is only indirectly converted into magnetic energy by stressing the magnetic field. 
Several experiments emulating different driver speeds on interacting fragments were carried out in \citet{Galsgaard1996,Galsgaard2000,Galsgaard2005} and showed that viscous dissipation was much smaller than  Joule dissipation. In all the experiments, the magnetic fragment was advected by the flow using an imposed speed. Yet here we have exposed the dual nature of the photospheric flows. In our observations, the velocity of the motion of the magnetic fragment can sometimes be much faster than the supergranular flows. Indeed, the approach speed $\dot{a}$ in Table~\ref{table3} can be up to $700\mets$, a typical value for magnetic elements in the quiet Sun \citep{Berger98}, whereas the mean velocity in supergranular flows and the large-scale vortex flows do not exceed a few hundred of $\mets$ \citep{Attie09}. See also the regions where the streamlines are more twisted: In \Fig{Vtransients} (E1 to E6), the streamlines and funnels are more sheared in flow fields of typically less than $400\mets$, with an exception maybe of E3 where the flows are sheared at a velocity of $\sim500-600\mets$ in the northern vicinity of the centre of the vortical streamlines. And yet again in \Fig{overview3} (bottom), the transients in C1, C2, D2, D1 are all in regions with flow speed of  less than $400\mets$. These velocities hardly match the faster approach speed ($\dot{a}$) of the magnetic features. Instead, there is a clear discrepancy between the average "supergranular" flows  and the motion of the photospheric cuts of magnetic flux tubes that we measured at a higher resolution. This brings additional useful information when considering the "advection" of magnetic fragments within the photospheric flows. This shall be taken into account when calculating the effect of shearing motions on magnetic stress, which otherwise could be underestimated.

Note however that the observed "faster" motion are deduced from observing the photospheric cut of magnetic flux tubes. We are aware that, from our perspective, the photospheric footpoints of an emerging or submerging magnetic loop would be observed as a horizontal flow \citep{Demoulin2003}.  With new tracking techniques like magnetic balltracking we are now able to track individual magnetic footpoints across the solar surface, and statistical analyses will help us to separate true horizontal flux tube motions from these "false" horizontal motions.

%How does the motion of magnetic flux tubes differ from the flows in which they supposedly "frozen"? What can this tell us on the local plasma-beta and pressure balance in the flux tubes that squeeze in the funnels and vortical flows? 

\subsection{Qualitative model of X-ray network flares}

Based on the present observations, we can describe the different steps that lead to the quiet Sun network flares that we have observed in the NFI FOV. Because the other network flares observed in the MDI FOV have the same characteristics (bipolar field underneath, located at the intersection of the supergranular lanes), this description may also apply to the transients E1 to E6 (\Fig{xrt_mdi1}).

\begin{enumerate}
\item Emergence phase: The magnetic flux emerges as small loops in the internetwork. 

\item Pre-interactive phase: The magnetic elements follow the funnelled streamlines, that are converging toward the vortex. The flux eventually clusters, merges again, and gets squeezed in as the funnels get tighter near the junction. In the presence of vortical flows, magnetic stress is increased and the energy eventually builds up. 

\item Energy release: The reconnection of the small core field with the overlying coronal field lines in "bald patches" frees plasma into the higher coronal loops, they are observed in X-ray as network flares. From the extrapolations, we cannot be sure of the true topology of the magnetic field; the plasma can be released either to larger coronal loops, or to the interplanetary magnetic field and populate the solar wind. 

\item Flux cancellation: As a result of magnetic dissipation and/or submergence enforced by the flows, the flux rapidly decreases. As the interaction distance is very small, the phase of energy release may overlap with the flux cancellation.
\end{enumerate}

From this we can conclude that the quiet Sun network flares require a specific flow pattern sketched in \Fig{cartoon2}. Funnels and vortices appear as the elementary flow structures that facilitate the compression of the magnetic elements, causing an increase of the flux density when the magnetic elements have the same polarity. They increase the probability of reconnection and subsequent cancellation when bipolar features are trapped in them. Funnels and vortices may be necessary, but not sufficient flow patterns to trigger the network flares. Thus one can anticipate the preferred (if not unique) sites of these localised soft X-ray emissions, whatever their actual nature is (micro-, nano-, or normal flares, or jets, micro-jets, mini CMEs, etc...), and for which we estimated the average energy flux to be two to three orders of magnitude less than required to heat the quiet Sun corona. This shall be investigated in future statistical studies.

\section{Prospects for future studies} 

In \citet{Graham08}, the higher resolution magnetograms from Hinode/SOT allowed a multi-scale study of the magnetic flux in the quiet Sun in which the self-similar pattern of the magnetic flux is quantified. This self-similar pattern holds for several orders of magnitude, including the small scales in which our study lies, and goes down to $20\,\mathrm{km}$ (i.e. below granular scales) using MHD simulations. 
In addition, we note that one common characteristic between granulation and supergranulation is that they sweep out, mix and disrupt the magnetic field at their respective boundaries. At granular scales, the flow is much faster than the supergranular flow, up to more than $1000\mets$ \citep{Berger98}. It contains a significant numbers of vortices, which have already been observed at smaller scale by \citet{Brandt88} and \citet{Bonet2010}. Thus, down-scaling the sketch in \Fig{cartoon2}, we can imagine that energy release within the smaller granular lanes also occurs, but at a faster rate and at smaller spatial scales as a result of the same interactions described in the network flares. 

In addition, EUV transient events were reported by \citet{Innes09} with similar scales (time and size) as the network flares. They are associated with propagating dim clouds, and/or propagating dim shock-waves, which makes them observationally equivalent to CMEs but at the scale of the network flares. The topology of the flows underneath also satisfied the necessary condition that we have assessed here, that is, the presence of vortical flow underneath the eruptions. Combined statistical studies of both EUV transient events and X-ray network flares is a key to better understand the dynamics of the quiet Sun, including their contribution to coronal heating and to solar wind acceleration. Such a survey is possible with the combined use of the Atmospheric Imaging Assembly (AIA) and the Helioseismic and Magnetic Imager (HMI) onboard the Solar Dynamic Observatory (SDO).

\citet{Wedemeyer2012} described chromospheric swirls resembling "magnetic tornadoes" as energy channels that reach the upper solar atmosphere, and it has been suggested that they are the result of rotating magnetic structures. Although these swirls are chromospheric structures, could the supergranular vortex flows be their photospheric trigger? To what extent are the funnels and the vortices reshaping the Sun's magnetic field topology? Could these flow patterns originate from the deeper layers of the solar atmosphere?
The use of tracking methods, such as the ones used in this work, may enable us to fill this knowledge gap.

\bibliographystyle{aa}
\bibliography{sun_v3}

\end{document}